\documentclass[aps,prd,groupedaddress,amssymb,twocolumn,eqsecnum,showpacs,epsfig,nofootinbib]{revtex4}
\usepackage{graphicx}
\usepackage{bm}
\usepackage{dcolumn}
\usepackage{amsmath}
\usepackage{amsmath}
\usepackage{amssymb}

\numberwithin{equation}{section}
\usepackage{epsfig}

\begin{document}
\newcommand{\newc}{\newcommand}

\newc{\be}{\begin{equation}}
\newc{\ee}{\end{equation}}
\newc{\bear}{\begin{eqnarray}}
\newc{\eear}{\end{eqnarray}}
\newc{\bea}{\begin{eqnarray*}}
\newc{\eea}{\end{eqnarray*}}
\newc{\D}{\partial}
\newc{\ie}{{\it i.e.} }
\newc{\eg}{{\it e.g.} }
\newc{\etc}{{\it etc.} }
{\newc{\etal}{{\it et al.}}
\newc{\lcdm}{$\Lambda$CDM }
\newcommand{\nn}{\nonumber}
\newc{\ra}{\rightarrow}
\newc{\lra}{\leftrightarrow}
\newc{\lsim}{\buildrel{<}\over{\sim}}
\newc{\gsim}{\buildrel{>}\over{\sim}}
\newcommand{\fs}{{\rm{\it f\sigma_8}}}
\newcommand{\mincir}{\raise
-3.truept\hbox{\rlap{\hbox{$\sim$}}\raise4.truept\hbox{$<$}\ }}
\newcommand{\magcir}{\raise
-3.truept\hbox{\rlap{\hbox{$\sim$}}\raise4.truept\hbox{$>$}\ }}

\title{Observational constraints on viable $f(R)$ parametrizations with geometrical and dynamical probes}

\author{Spyros Basilakos}\email{svasil@academyofathens.gr}
\affiliation{Academy of Athens, Research Center for Astronomy and
Applied Mathematics,
 Soranou Efesiou 4, 11527, Athens, Greece}

\author{Savvas Nesseris}\email{nesseris@nbi.ku.dk}
\affiliation{Instituto de F\'isica Te\'orica UAM-CSIC, Universidad Auton\'oma de Madrid,
Cantoblanco, 28049 Madrid, Spain}

\author{Leandros Perivolaropoulos}\email{leandros@uoi.gr}
\affiliation{Department of Physics, University of Ioannina, Greece}

\date{\today}

\begin{abstract}
We demonstrate that a wide range of viable $f(R)$ parameterizations (including the Hu \& Sawicki and the Starobinsky models) can be expressed as perturbations deviating from the \lcdm Lagrangian. We constrain the deviation parameter $b$ using a combination of geometrical and dynamical observational probes. In particular, we perform a joint likelihood analysis of the recent Supernovae type Ia data, the Cosmic Microwave Background shift parameters, the Baryonic Acoustic Oscillations and the growth rate data provided by the various galaxy surveys. This analysis provides constraints for the following parameters: the matter density $\Omega_{m0}$, the deviation from \lcdm parameter $b$ and the growth index $\gamma(z)$.  We parametrize the growth index $\gamma(z)$ in three manners (constant, Taylor expansion around $z=0$, and Taylor expansion around the scale factor). We point out the numerical difficulty for solving the generalized $f(R)$ Friedman equation at high redshifts due to stiffness of the resulting ordinary differential equation. We resolve this problem by constructing an efficient analytical perturbative method in the deviation parameter $b$. We demonstrate that this method is highly accurate, by comparing the resulting analytical expressions for the Hubble parameter, with the numerical solutions at low and intermediate redshifts. Surprisingly, despite of its perturbative nature, the accuracy of the method persists even for values of $b$ that are of $O(1)$.

\end{abstract}
\maketitle

\section{Introduction}
A variety of cosmological studies
have converged to a cosmic expansion history involving a spatially
flat geometry and a cosmic dark sector formed by cold dark matter and some
sort of dark energy, endowed with large negative pressure, in order to
explain the observed accelerating expansion of
the Universe\cite{Teg04,Spergel07,essence,Kowal08,Hic09,komatsu08,LJC09,BasPli10,komatsu11}.
In this framework, the absence of a fundamental physical theory, regarding
the mechanism inducing the cosmic acceleration, has given rise to a
plethora of alternative cosmological scenarios.
Modified gravity models act
as an important alternative to
the scalar-field dark energy models, since
they provide an efficient way towards explaining
the accelerated expansion of the universe, under a modification of
the nature of gravity.
Such an approach is an attempt to evade the
coincidence and cosmological constant problems of
the standard $\Lambda$CDM model.

Particular attention over the last decades has been paid on
 $f(R)$ gravity theories \cite{Sot10}.
In this scenario of nonstandard
gravity, one modifies the Einstein-Hilbert action with a general
function $f(R)$ of the Ricci scalar $R$. The $f(R)$ approach is a relative
simple but fundamental tool used to explain the accelerated expansion
of the universe. A pioneering approach was proposed long ago,
where $f(R)=R+mR^{2}$\thinspace\ \cite{Star80}. Later on, the $f(R)$ models
were further explored from different points of view in
\cite{Carrol,FR,Amendola-2007a} and a large number of functional forms
of $f(R)$ gravity is currently available in the literature. It is interesting
to mention that subsequent investigations \cite{Amendola-2007a} confirmed
that $1/R$ gravity is an unacceptable model because it fails to reproduce the
correct cosmic expansion in the matter era.
Of course, there are many other possibilities to explain the present
accelerating stage.
Indeed, in the literature one can find a
variety of modified
gravity models (for reviews see Refs. \cite{Ame10})
which include the braneworld
Dvali, Gabadadze and Porrati (hereafter DGP; \cite{DGP}) model,
Finsler-Randers gravity \cite{Stav12}, scalar-tensor theories \cite{scal}
and Gauss-Bonnet gravity \cite{gauss}.

The construction of observationally viable $f(R)$ theories has proved to be challenging because it has been shown \cite{Amendola-2007a}  that most of these models do not predict a matter era in the cosmic expansion history. Nevertheless observationally viable $f(R)$ models have been constructed and two such examples are the following:
\begin{enumerate}
\item
The Hu \& Sawicki model \cite{Hu07} with
\begin{equation}
\label{Hu}
f(R)=R-m^2 \frac{c_1 (R/m^2)^n}{1+c_2 (R/m^2)^n}
\end{equation}
where $c_1$, $c_2$ are free parameters, $m^2\simeq \Omega_{m0}H^{2}_{0}$ is of the
order of the Ricci scalar $R_{0}$ at
the present time, $H_{0}$ is the Hubble constant, $\Omega_{m0}$ is the
dimensionless matter density parameter at the present time,
$m$ and $n$ are positive constants.
\item
The Starobinsky model \cite{Starobinsky-2007} with
\begin{equation}
\label{Star}
f(R)=R-c_1~m^2 \left[1-\left(1+R^2/m^{4}\right)^{-n}\right]\;.
\end{equation}
\end{enumerate}
 These models were originally advertised as models that do not contain the cosmological constant as part of $f(R)$ being distinct from the \lcdm form $f(R)= R - 2 \Lambda $ (where $\Lambda$ is the cosmological constant). However, it is straightforward to show that both the Hu \& Sawicki and the Starobinsky models may be written in terms of \lcdm modified by a distortion function $y(R,b)$ that depends on a deviation parameter $b$ as:
\begin{equation}
\label{mod-lcdm}
f(R)=R-2\Lambda\; y(R,b)
\end{equation}
where\cite{Bamba:2012qi}
\begin{equation}
\label{hsgr}
y(R,b)=1-\frac{1}{1+(R/(b~\Lambda)^n}
\end{equation}
for the Hu \& Sawicki model with $\Lambda= \frac{m^2 c_1}{2c_2}$ and $b=\frac{2 c_2^{1-1/n}}{c_1}$, while
\begin{equation}
\label{stargr}
y(R,b)=1- \frac{1}{\left(1+\left(\frac{R}{b \Lambda }\right)^2\right)^n}
\end{equation}
for the Starobinsky model where $\Lambda= \frac{c_1 m^2}{2}$ and $b=\frac{2}{c_1}$. Notice that in both cases
the following two limits exist for $n>0$:
\bear
\lim_{b\rightarrow0}f(R)&=&R-2\Lambda \nn \\
\lim_{b\rightarrow \infty}f(R)&=&R
\eear
and therefore both models reduce to \lcdm for $b\rightarrow 0$. Notice that both the Hu \& Sawicki and the Starobinsky models effectively include the cosmological constant even though they were advertised as being free from a cosmological constant in the original papers \cite{Hu07,Starobinsky-2007}. In fact by proper choices of the function $y(R,b)$ it is possible to construct infinite viable $f(R)$ models which however will always include \lcdm as a limiting case for $b\rightarrow 0$.

Thus, an important question that arises is the following: {\it `What is the range of the deviation parameter $b$ that is consistent with cosmological observations?'} This is the main question addressed in the present study. Since \lcdm is consistent with observations we anticipate that the value $b=0$ is within the acceptable range of $b$ values. Thus, the interesting part of the question is {\it `What is the maximum allowed value of $b$ at eg the $2\sigma$ confidence level?'}

In order to address this question we solve the background modified Friedman equation assuming flatness and obtain the Hubble parameter $H(\Omega_{m0},b;z)$. This involves the numerical solution of a stiff ordinary differential equation (ODE), second order in $H$, with initial conditions at high $z$ that correspond to \lcdm. The numerical solution of this stiff ODE at redshifts higher than $z\simeq 300$ is quite challenging. However, we have developed an efficient analytical perturbative expansion in $b$ to solve it. This expansion leads to an analytic expression for $H(\Omega_{m0},b;z)$ to all orders in $b$. We thus use geometric probes (type Ia surenovae, CMB shift parameter and Baryon Acoustic Oscillation data) to constrain the parameters $\Omega_{m0},b$ that appear in the expression of $H(\Omega_{m0},b;z)$.

In addition to geometric observations that probe directly the cosmic metric, dynamical probes play a crucial role in constraining cosmological models. The growth index, $\gamma$, could provide an efficient way to discriminate between modified gravity models and
scalar field dark energy (hereafter DE) models which
adhere to general relativity.
The accurate determination of the growth index is considered one of
the most fundamental tasks for Observational
Cosmology. Its importance steams from the fact that
there is only a weak dependence of $\gamma$ on
the equation of state parameter $w(z)$, as has been found by
Linder \& Cahn \cite{Linder2007}, which implies that one can separate the background expansion history, $H(z)$, constrained by geometric probes (SnIa, BAO, CMB), from the fluctuation growth history, given by $\gamma$.
For a constant DE equation of state $w$, it was theoretically shown that for DE models
within general relativity the growth index $\gamma$ is
well approximated by $\gamma \simeq \frac{3(w-1)}{6w-5}$
(see Refs. \cite{Linder2007},\cite{Silv94},\cite{Wang98},\cite{Nes08}), which
boils down to $\approx 6/11$ for the $\Lambda$CDM cosmology $w(z)=-1$.
Notice, that in the case of the
braneworld model of Dvali, Gabadadze \& Porrati \cite{DGP}
we have $\gamma \approx 11/16$
(see also Refs. \cite{Linder2007,Gong10,Wei08,Fu09}), while for some f(R) gravity models we have $\gamma \simeq 0.415-0.21z$
for various parameter values (see \cite{Gann09,Tsu09}).
Recently, Basilakos \& Stavrinos \cite{Bastav13}
found $\gamma \approx 9/14$ for the Finsler-Randers cosmology.

Observationally, indirect methods
to measure $\gamma$ have also been developed (mostly using a
constant $\gamma$),
based either on the observed growth rate of
clustering \cite{Nes08,Guzzo08,Port08,Gong10,Dos10,Por,Hud12,Bass,Samnew12}
providing a wide range of $\gamma$ values
$\gamma\simeq (0.58-0.67)$,
or on massive galaxy clusters (Vikhlinin et al. \cite{Vik09} and
Rapetti et al. \cite{Rap10}). The
latter study provides $\gamma=0.42^{+0.20}_{-0.16}$.
An alternative method for measuring $\gamma$ involves weak gravitational lensing \cite{Daniel10}.
Gaztanaga et al. \cite{Gazt12} performed
a cross-correlation analysis between probes of weak gravitational lensing and
redshift space distortions and found no evidence for deviations
from general relativity.
Also, Basilakos \& Pouri \cite{Por} and
Hudson \& Turnbull \cite{Hud12} imposed constraints on the growth index using
the combination parameter
$F(z)\sigma_{8}(z)$ \footnote{Here the capital
$F(a)$ denotes the growth rate of structure. We follow the latter notation in order
to avoid confusion with the $f(R)$.}, of the growth rate of structure
$F(z)$ multiplied by the redshift-dependent rms fluctuations of the linear
density field, $\sigma_8(z)$.
The above authors found $\gamma=0.602\pm 0.05$ \cite{Por} $\gamma=0.619\pm 0.05$ \cite{Hud12} while
Basilakos \& Pouri \cite{Por} showed
that the current growth data can not accommodate
the Dvali, Gabadadze \& Porrati \cite{DGP} gravity model.

In order to impose constraints on the viable $f(R)$ models discussed above, we use
in addition to geometric probes, the recent growth rate data
as collected by Nesseris \& Garcia-Bellido \cite{NesGa12}, Hudson
\& Turnbull \cite{Hud12}
and Beutler et al. \cite{Beutler}.

The plan of the paper is as follows.
Initially in section II, we briefly discuss the background
cosmological equations. The basic features of
the growth index are presented in section III, where
we extend the original Polarski \& Gannouji
method \cite{Pol} for a general family of $\gamma(z)$
parametrizations as well as $f(R)$ cosmological models.
In section~IV, a joint statistical analysis
based on the {\em Union 2.1} set of type Ia supernovae
(SnIa \cite{Suzuki:2011hu}), the observed Baryonic
Acoustic Oscillations (BAOs \cite{Perc10}),
the shift parameter of the Cosmic
Microwave Background (CMB \cite{komatsu11}), and the
observed linear growth rate of clustering, measured
mainly from the PSCz, 2dF, VVDS, SDSS, 6dF, 2MASS, BOSS and {\em WiggleZ}
redshifts catalogs,
is used to constraint the growth index model free parameters.
Finally, we summarize our main conclusions in section V.

\section{The background evolution}
First of all we start with the assumption that
the universe is a
self-gravitating fluid described by a modified gravity namely $f(R)$
\cite{Sot10}, and endowed with a spatially flat homogeneous and
isotropic geometry. In addition, we also consider that the
universe is filled by non-relativistic
matter and radiation. The modified Einstein-Hilbert action
reads:

\begin{equation}
S=\int d^{4}x\sqrt{-g}\left[  \frac{1}{2k^{2}}f\left(  R\right)
+\mathcal{L}_{m}+\mathcal{L}_{r}\right]  \label{action1}%
\end{equation}
where $\mathcal{L}_{m}$
is the Lagrangian of matter,
$\mathcal{L}_{r}$ is the Lagrangian of radiation and
$k^{2}=8\pi G$. Now varying the action with respect to the metric\footnote{We
use the metric i.e. the Hilbert variational approach.} we arrive at
\begin{align}
&  f_{R} \; G_{\nu}^{\mu}\,-\,g^{\mu\alpha}f_{_{R},\,\alpha\,;\,\nu
}+\left[  \frac{2\Box f_{R}-(f-Rf_{R})}{2}\right]
\delta_{\;\nu}^{\mu}\nonumber\\
&  =k^{2}\,T_{\nu}^{\mu} \;.
\label{EE}%
\end{align}
where $R$ is the Ricci scalar, $f_{R}=\partial f/\partial R$, $G_{\nu}^{\mu}$
is the Einstein tensor and $T_{\nu}^{\mu}$ is the energy-momentum tensor of
matter.
Modeling the expanding universe as a perfect fluid
that includes radiation and cold dark matter
with $4-$velocity $U_{\mu}$, we have $T_{\nu}^{\mu}=-P\,g_{\nu}^{\mu}+
(\rho+P)U^{\mu}U_{\nu}$, where $\rho=\rho_{m}+\rho_{r}$ and $P=p_{m}%
+p_{r}$ are the total energy density and pressure of the cosmic fluid
respectively. Note that $\rho_{m}$ is the matter density, $\rho_{r}$
denotes the density of the radiation
and $p_{m}=0$, $p_{r}=\rho_{r}/3$ are the corresponding pressures.
Assuming negligible interaction between non-relativistic matter and radiation
the Bianchi identity $\bigtriangledown^{\mu}\,{T}_{\mu\nu}=0$
(which insures the covariance of the theory) leads to the matter/radiation
conservation laws:
\begin{equation}
\dot{\rho}_{m}+3H\rho_{m}=0 \;\;\;\;\;\dot{\rho}_{r}+4H\rho_{r}=0
\, \label{frie3}%
\end{equation}
the solutions of which are
$\rho_{m}=\rho_{m0}a^{-3}$ and $\rho_{r}=\rho_{r0}a^{-4}$.
Note that the over-dot denotes derivative with respect to the cosmic time $t$, $a(t)$ is the scale factor and $H\equiv\dot{a}/a$ is the Hubble parameter.

Now, in the context of a flat FLRW metric with Cartesian coordinates
\begin{equation}
ds^{2}=-dt^{2}+a^{2}(t) (dx^{2}+dy^{2}+dz^{2}) \label{SF.1}%
\end{equation}
the Einstein's tensor components are given by:
\begin{equation}
\label{EIN.1}
G_{0}^{0}=-3H^{2}, \;\;\;\; G^{i}_{i}=-\delta^{\mu}_{\nu
}\left(  2\dot{H}+3H^{2}\right)  \;.
\end{equation}
Inserting Eqs.(\ref{EIN.1}) into the modified Einstein's field equations
(\ref{EE}), for comoving observers, we derive the modified Friedmann's
equations
\begin{equation}
3f_{R}H^{2}-\frac{f_{R}R-f}{2}+3Hf_{RR}\dot{R} =
k^{2}(\rho_{m}+\rho_{r})
\label{motion1}%
\end{equation}
\begin{equation}
-2f_{R}\dot{H}=k^{2}[\rho_{m}+(4/3)\rho_{r}] +\ddot{f}_{R}-H\dot{f}_{R}
\label{motion2}%
\end{equation}
where $\dot{R}=aHdR/da$ and $f_{RR}=\partial^{2} f/\partial R^{2}$.
Also, the contraction of the Ricci tensor provides the Ricci scalar
\begin{equation}
\label{SF.3b}
R=g^{\mu\nu}R_{\mu\nu}= 6\left(  \frac{\ddot{a}}{a}+\frac{\dot
{a}^{2}}{a^{2}}\right)  =6(2H^{2}+\dot{H}) \;.
\end{equation}

Of course, if we consider $f(R)=R$ then
the field equations (\ref{EE}) boil down to the nominal Einstein's equations a solution of which is the Einstein de Sitter model. On the other hand, the concordance $\Lambda$ cosmology is fully recovered for $f(R)=R-2\Lambda$.
We would like to stress here that within the context of the metric formalism the above
$f(R)$ cosmological models must obey simultaneously some strong conditions
(for an overall discussion see \cite{Ame10}). Briefly these are: (i)
$f_R>0$ for $R\ge R_{0}>0$, where $R_{0}$ is the Ricci scalar at the
present time. If the final attractor is a de Sitter point we need to have
$f_R>0$ for $R\ge R_{1}>0$, where $R_{1}$ is the Ricci scalar at the
de Sitter point, (ii) $f_{RR}>0$ for $R\ge R_{0}>0$, (iii) $f(R)
\approx R-2\Lambda$ for $R\gg R_{0}$ and finally (iv) $0<\frac{Rf_{RR}}{f_R}(r)<1$ at $r=-\frac{Rf_R}{f}=-2$.

Finally, from the current analysis it becomes clear that unlike the standard Friedmann
equations in Einstein's general relativity the modified equations of motion (\ref{motion1}) and (\ref{motion2}) are complicated and thus it is
difficult to solve them
analytically. Below, we are going to compare
the $f(R)$ results with those of the
concordance $\Lambda$CDM model. This can help us to
understand better the theoretical basis of the current
$f(R)$ models as well as the variants from general relativity.
For practical reasons (see below), we derive the
the effective (``geometrical'') dark energy
EoS parameter in terms of $E(a)=H(a)/H_{0}$ (see \cite{Ame10,Saini00}
and references therein)
\begin{equation}
\label{eos222}
w(a)=\frac{-1-\frac{2}{3}a\frac{{d\rm lnE}}{da}}
{1-\Omega_{m}(a)}
\end{equation}
where
\be
\label{ddomm}
\Omega_{m}(a)=\frac{\Omega_{m0}a^{-3}}{E^{2}(a)} \;.
\ee
Differentiating the latter and utilizing
Eq.~(\ref{eos222}) we find that
\be
\label{domm}
\frac{d\Omega_{m}}{da}=
\frac{3}{a}w(a)\Omega_{m}(a)\left[1-\Omega_{m}(a)\right]\;.
\ee
In the case of the traditional $\Lambda$CDM cosmology $f(R)=R-2\Lambda$, the
corresponding dark energy EoS parameter is strictly equal to $-1$ and the normalized
Hubble function in the matter era is given by
\be
E_{\Lambda}(a)=\left( \Omega_{m0}a^{-3}+1-\Omega_{m0}\right)^{1/2} \;\;.
\ee

\subsection{The $f(R)$ functional forms}
In order to solve numerically
the modified Friedmann equation (\ref{motion1}) we
need to know apriori the functional form of $f(R)$. Due to the absence of a
physically well-motivated functional form for the $f(R)$ parameter, there are
many theoretical propositions in the literature.
In this article for the background we use different reference expansion models
namely flat $\Lambda$CDM and $f(R)$ respectively. Bellow we briefly present
the two most popular $f(R)$ models whose free parameters can be constrained from the current cosmological data.

Firstly, we use the Hu \& Sawicki \cite{Hu07} model (hereafter $f_{1}$CDM) as expressed by equation (\ref{Hu}).
Using the constrains provided by the violations of weak and strong
equivalence principle, Capozziello \& Tsujikawa \cite{CapTs}
found that $n>0.9$. On the other hand it has been proposed by \cite{FuWu}
that $n$ is an integer number, so for simplicity in our
work we have set $n=1$. In Ref.~\cite{Hu07} the parameters $(c_1,c_2)$ were related to $\Omega_{m0}$, $\Omega_{r0}$ and the
first derivative of $f(R)$ at the present epoch $f_{R0}$ in order to ensure the expansion history is close to that of $\Lambda$CDM. Specifically, we have
\bea
\frac{c_1}{c_2}&=&6\frac{ \left(1-\Omega_{r0}-
\Omega_{m0}\right)}{\Omega_{m0}} \\
f_{R0}&=&1-\frac{n c_1}{c_2^2} \left(-9+\frac{12}{\Omega_{m0}}-\frac{12
\Omega_{r0}}{\Omega_{m0}}\right)^{-1-n}
\eea

The first two derivatives of Eq.(\ref{Hu}) with respect to $R$ are
\begin{equation}
\label{dHu1}
f_{R}=\frac{R \left[c_2 \left(\frac{R}{m^2}\right)^n+1\right]^{2}-c_1 m^2 n \left(\frac{R}{m^2}\right)^n}{R \left[c_2 \left(\frac{R}{m^2}\right)^n+1\right]{^2}}
\end{equation}

\begin{equation}
\label{dHu2}
f_{RR}=\frac{c_1 m^2 n \left(\frac{R}{m^2}\right)^n \left[c_2 (n+1) \left(\frac{R}{m^2}\right)^n-n+1\right]}{R^2 \left[c_2 \left(\frac{R}{m^2}\right)^n+1\right]{^3}}
\end{equation}
As discussed in the Introduction, the Lagrangian of Eq.~(\ref{Hu}) can also be written as
\bear
\label{Hu1}
f(R)&=& R- \frac{m^2 c_1}{c_2}+\frac{m^2 c_1/c_2}{1+c_2 (R/m^2)^n} \nn\\
&=& R- 2\Lambda\left(1-\frac{1}{1+(R/(b~\Lambda)^n}\right) \nn \\
&=& R- \frac{2\Lambda }{1+\left(\frac{b \Lambda }{R}\right)^n}
\eear where $\Lambda= \frac{m^2 c_1}{2c_2}$ and $b=\frac{2 c_2^{1-1/n}}{c_1}$. In this form it is clear that the HS model can be arbitrarily close to $\Lambda$CDM, depending on the parameters $b$ and $n$.

We now consider the Starobinsky \cite{Starobinsky-2007} model (hereafter $f_{2}$CDM) as expressed by equation (\ref{Star}):
As in the Hu \& Sawicki \cite{Hu07} model we choose $m^2\simeq
\Omega_{m0}H^{2}_{0}$, while \cite{CapTs} also showed that $n>0.9$.
In this case the $f_{R}$ and $f_{RR}$ derivatives are given by
\begin{equation}
\label{dsta1}
f_{R}=1-\frac{2 n R \left(1+\frac{R^2}{m^4}\right)^{-1-n} c_1}{m^2}
\end{equation}

\begin{equation}
\label{dsta2}
f_{RR}=-\frac{2 m^2 n \left(1+\frac{R^2}{m^4}\right)^{-n} \left[m^4-(1+2 n) R^2\right] c_1}{\left(m^4+R^2\right)^2}
\end{equation}
In order to ensure that the expansion history of this model is close to that of $\Lambda$CDM we need to match the $c_1$ constant to $\Lambda$, ie
$-c_1~m^2=-2\Lambda=-6(1-\Omega_{m0}-\Omega_{r0}) H_0^2$ or
\be
c_1=\frac{6 (1-\Omega_{m0}-\Omega_{r0})}{\Omega_{m0}}
\ee As discussed in the Introduction, the Lagrangian of Eq.~(\ref{Star}) can also be written as
\be
\label{Star1}
f(R)=R-2\Lambda  \left[1- \frac{1}{\left(1+\left(\frac{R}{b \Lambda }\right)^2\right)^n}\right]
\ee where $\Lambda= \frac{c_1 m^2}{2}$ and $b=\frac{2}{c_1}$. In this form it is clear that that this model can also be arbitrarily close to $\Lambda$CDM, depending on the parameters $b$ and $n$. Thus, the parameter $b$ determines how close the model is to $\Lambda$CDM.

It is interesting to mention that the above $f(R)$ models
satisfy all the strong
conditions (see section II) and thus they provide predictions
which are similar to those of the usual (scalar field)
DE models, as far as the cosmic history (presence of
the matter era, stability of cosmological perturbations, stability
of the late de Sitter point etc.) is concerned. Also, we will
restrict our present numerical solutions to the
choice $H_{0}=70.4$Km/s/Mpc  and $\sigma_8=0.8$. For example, in this
case the modified Friedmann equations for the $f(R)$ models
contain two free parameters, namely $(\Omega_{m0},b)$
which can be constrained from the current
cosmological data.

\subsection{Analytic approximations}
In this subsection we present a novel approximation scheme for the solution of the modified Friedmann equation (\ref{motion1}) and we explicitly apply it to the two widely used models Eqs.~(\ref{Hu1}) and (\ref{Star1}).

In particular we may write eq. (\ref{motion1}) as
\bear
\label{motionn}
&&-f_R H^2(N)+(\Omega_{m0}e^{-3N}+\Omega_{r0} e^{-4N})+\frac{1}{6}(f_R R-f)\; \nonumber \\
&&=f_{RR} H^2(N) R'(N) \;,
\eear
where the prime denotes differentiation with respect to $N$ and $R(N)$ is given by eq. (\ref{SF.3b}). Using now eq. (\ref{mod-lcdm}) for specific $f(R)$ models, the above ODE (and its solution $H(N)$) may be expanded around \lcdm with respect to the deviation parameter $b$.

Since we are interested in testing deviations from the $\Lambda$CDM model, we find it useful to perform a series expansion of the solution of the ODE (\ref{motionn}) around $b=0$ as
\be
H^2(N)=H_{\Lambda}^2(N)+\sum_{i=1}^M b^i \delta H_i^2(N) \label{expansion1}
\ee
where
\be
\frac{H_{\Lambda}^2(N)}{H_0^2}=\Omega_{m0}e^{-3N}+\Omega_{r0} e^{-4N}+(1-\Omega_{m0}-\Omega_{r0})\label{LCDM1}
\ee and $M$ is the number of terms we keep before truncating the series. Usually  keeping only the two first non-zero terms is more that enough to have excellent agreement of \textit{better than} $0.001\%$ at all redshifts with the numerical solution for realistic values of the parameter $b\in[0.001,0.5]$.

\begin{figure*}[t!]
\centering
\vspace{0cm}\rotatebox{0}{\vspace{0cm}\hspace{0cm}\resizebox{0.48\textwidth}{!}{\includegraphics{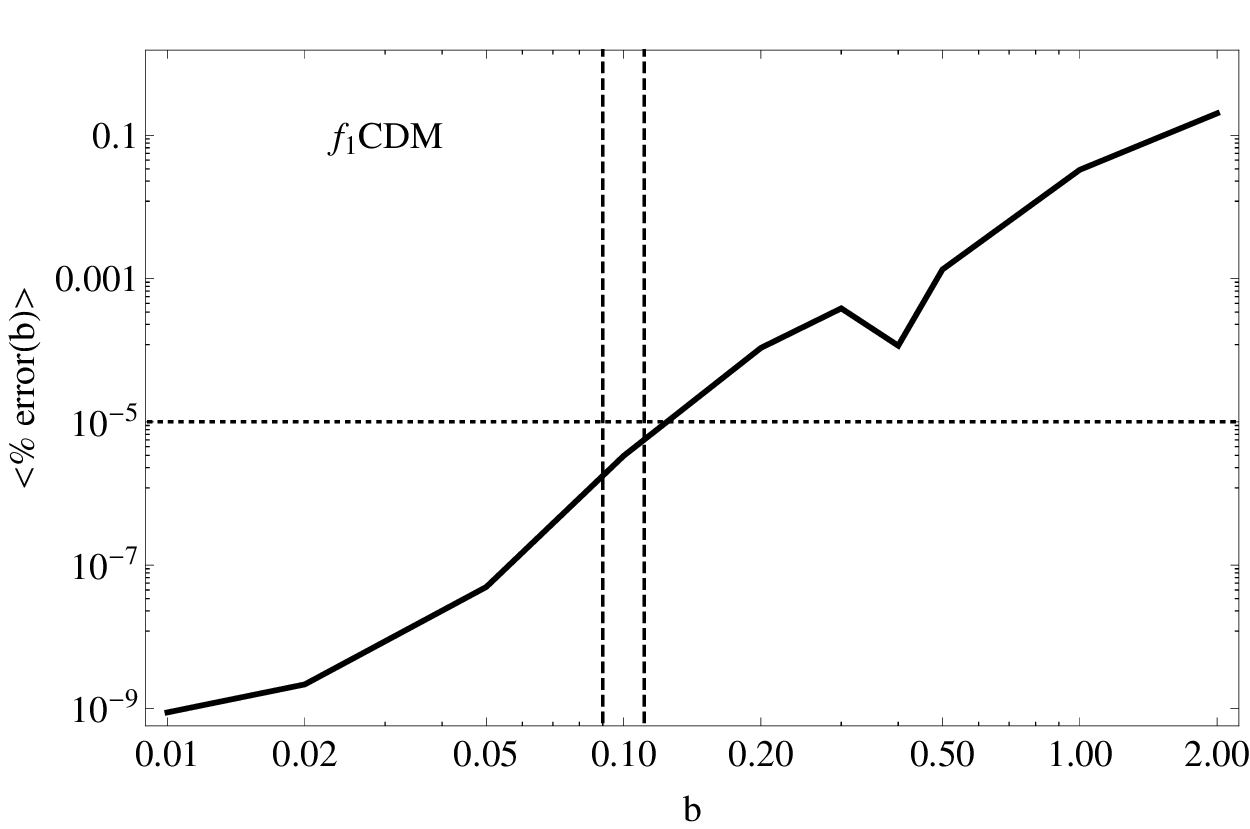}}}
\vspace{0cm}\rotatebox{0}{\vspace{0cm}\hspace{0cm}\resizebox{0.48\textwidth}{!}{\includegraphics{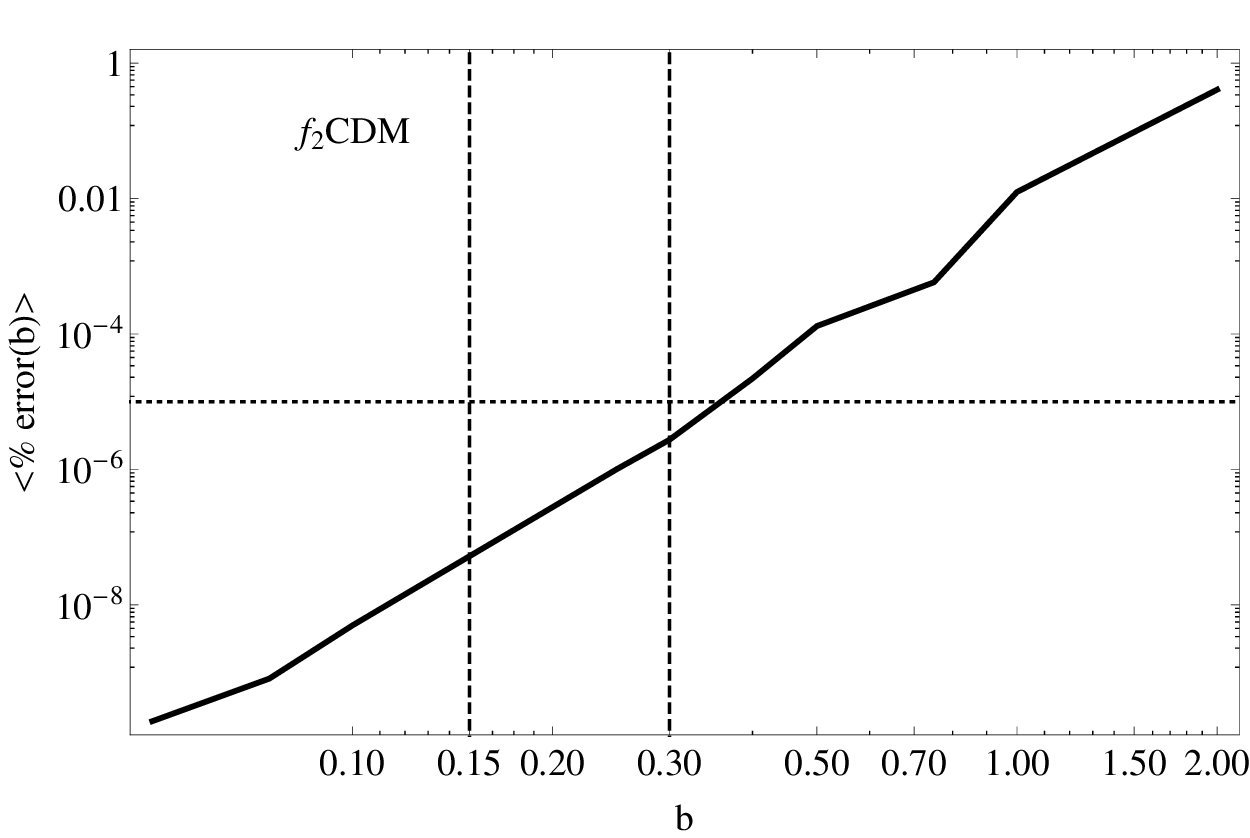}}}
\caption{The results of the average error in the redshift range $z\in[0,30]$ for a large variety of values of the parameter $b\in[0.01,2]$. Clearly, the approximation behaves exceptionally especially within the ranges of interest, ie the vertical dashed lines. The horizontal dotted line indicates an error of $10^{-5}\%$.\label{accaprox1}}
\end{figure*}

By expanding Eq.~(\ref{motionn})  with  (\ref{expansion1}) to any given order in $b$ we can find analytical solutions for the Hubble expansion rate. It is easy to show that for the HS model and for $n=1$ the first two terms of the expansion are the following:

\be
H_{HS}^2(N)=H_{\Lambda}^2(N)+b~\delta H_1^2(N)+ b^2~\delta H_2^2(N)+...\label{expansionHS1}
\ee
where $\delta H_1^2(N)$ and $\delta H_2^2(N)$ are given by Eqs.~(\ref{HSap1}) and (\ref{HSap2}) respectively. For the Starobinsky model for $n=1$ we have

\be
H_{Star}^2(N)=H_{\Lambda}^2(N)+b^2~\delta H_2^2(N)+ b^4~\delta H_4^2(N)+...\label{expansionstar1}
\ee
where $\delta H_2^2(N)$ and $\delta H_4^2(N)$ are given by Eqs.~(\ref{Starap1}) and (\ref{Starap2}) respectively.

Obviously, similar expressions can be obtained for any $f(R)$ model and up to any order provided that for $b\rightarrow 0$ we obtain \lcdm. We stress that the expressions for $\delta H_i^2(N)$ are algebraic up to all orders, something that makes this method very useful and fast compared to solving the differential equation numerically. Furthermore, this method avoids another problem of the numerical integration, namely that at very high redshifts the ODE of Eq.~(\ref{motion1}) is quite stiff, thus making the integration impossible with standard methods. This makes the numeric solution quite time consuming and possibly unreliable, something which as we will show is not a problem for our analytic approximation.

In what follows we will test the validity of this approximation in both cases. In order to do this we compared the predictions of the analytical solutions of Eqs.~(\ref{expansionHS1}) and (\ref{expansionstar1}) to the numerical solution in each case for a large variety of values for the parameter $b$. In particular, we estimated the average percent deviation between the approximations and the numerical solution, defined as:
\be
\left<\textrm{error}(b)\right>~=~\left<100\cdot\left(1-\frac{H_{approx}(z,b)}{H_{numeric}(z,b)}\right)\right>
\ee
where the average is taken over redshifts in the range $z\in[0,30]$. The reason for averaging is that most of the data we will use involve distance scales, like the luminosity distance, that are integrals of the Hubble parameter. We have also kept $z$ below $200$ since the numerical ODE solver of Mathematica is unable to go to larger redshifts due to the stiffness of the ODE.

We show the results of the average error for a large variety of values of the parameter $b\in[0.01,2]$ in Fig. \ref{accaprox1}. Clearly, the approximation behaves exceptionally especially within the ranges of interest, ie the vertical dashed lines. These regions correspond, as we will see in a later section, to the best fit values of $b$. The horizontal dotted line indicates an error of $10^{-5}\%$. Finally, we see that our approximation is on average in agreement to better than $\sim0.01\%$ for realistic parameters, ie $b\sim O(1)$, of the $f(R)$ models. In a forthcoming paper we attempt to investigate the validity of our method against all the available $f(R)$ gravity models.

We should note that there is some similarity between the
iterative approach suggested by Starobinsky
in Ref. \cite{Starobinsky:2007hu} and our method. Both approaches
are based on taking small deviations from an unperturbed simple case.
However, our approach is based on the existence of a well defined
dimensionless deviation parameter $b$ while Starobinsky uses the WKB approximation
with no reference to a deviation perturbative parameter.

In addition, in Ref.~\cite{Starobinsky:2007hu} the iterative procedure has been based on
the assumption that the Ricci scalar can be written in terms of three
components namely $R^{(0)}$, $\delta R_{ind}$ and $\delta R_{osc}$
(see Starobinsky's eq.12), whereas in our approach we perform a Taylor expansion of the Hubble function
around $b=0$. The reason of using such an expansion is due to the fact that for $b$ close to zero both $f(R)$
models tend to the concordance $\Lambda$CDM model.

Our approach is indeed a perturbative approach and it should be applicable for small values of the deviation
parameter $b$. The fact that the method remains accurate even for values of $b$ of O(1) can be attributed to the
fact that even for $b$ of $O(1)$ the deviation term as a whole remains small. As expected however, the accuracy of
the method decreases for increasing $b$ (see Fig. 1). We will compare
the above iterative procedures in a forthcoming paper.

\section{The Evolution of the linear growth factor}\label{sec:growth}
In this section we concentrate on the sub-horizon scales in which the DE component
is expected to be homogeneous and thus we can use
perturbations only on the matter component of the
cosmic fluid \cite{Dave02}. Therefore,
the evolution equation of the
matter fluctuations, for cosmological models
where the DE fluid has a vanishing anisotropic
stress and the matter fluid is not coupled to other species
(see \cite{Gann09},\cite{Lue04},\cite{Linder05},\cite{Stab06},\cite{Uzan07},\cite{Tsu08},\cite{Dent}), is given by:
\be
\label{odedelta}
\ddot{\delta}_{m}+ 2H\dot{\delta}_{m}=4 \pi G_{\rm eff} \rho_{m} \delta_{m}
\ee
where $\rho_{m}$ is the matter density and $G_{\rm eff}(t)=G_{N} Q(t)$, with $G_{N}$ denoting Newton's gravitational constant.

For those cosmological models which adhere to general relativity,
[$Q(t)=1$, $G_{\rm eff}=G_{N}$], the above equation reduces to the usual time evolution equation for the mass density contrast \cite{Peeb93}, while in the
case of modified gravity models (see \cite{Lue04},\cite{Linder2007},
\cite{Tsu08},\cite{Gann09}), we have $G_{\rm eff}\ne G_{N}$ (or $Q \ne 1$).
Indeed it has been shown (see Refs. \cite{Gann09,Tsu09}) that
in the case of $f(R)$ models the quantity $Q$ is a function of the
scale factor and of the wave-number $\kappa$
\be
\label{Newton}
Q(a,k)=\frac{1}{f_{R}}\frac{1+4\frac{\kappa^{2}}{a^{2}} \frac{f_{RR}}{f_{R}}}
{1+3\frac{\kappa^{2}}{a^{2}} \frac{f_{RR}}{f_{R}}} \;.
\ee
We restrict our analysis to the choice of
$\kappa=1/\lambda=0.1h$Mpc$^{-1}$ or $\lambda=10h^{-1}$Mpc
(see also \cite{Zhang12}).

In this context,
$\delta_{m}(t) \propto D(t)$, where $D(t)$ is the linear growing mode
(usually scaled to unity at the present time).
Of course, solving Eq.(\ref{odedelta}) for the
concordance $\Lambda$ cosmology, we
derive the well known perturbation growth factor (see \cite{Peeb93}):
\be\label{eq24}
D_{\Lambda}(z)=\frac{5\Omega_{m0}
  E_{\Lambda}(z)}{2}\int^{+\infty}_{z}
\frac{(1+u)du}{E^{3}_{\Lambda}(u)} \;\;.
\ee
In this work we use the above equation normalized to unity at the
present time.

Since in most of the cases Eq.(\ref{odedelta}) does not yield
analytical solutions,
it is common in this kind of studies to provide
an efficient parametrization
of the matter perturbations which is based on the
growth rate of clustering
\cite{Peeb93}
\be
\label{fzz221}
F(a)=\frac{d\ln \delta_{m}}{d\ln a}\simeq \Omega^{\gamma}_{m}(a)
\ee
where $\gamma$ is the growth index
(see Refs.~\cite{Silv94,Wang98,Lue04,Linder2007,Nes08})
which plays a key role in cosmological studies as we described in the
introduction.

\subsection{The generalized growth index parametrization}
Inserting the first equality of
Eq.(\ref{fzz221}) into Eq.~(\ref{odedelta})
and using simultaneously
Eq.~(\ref{eos222}) and $\frac{d}{dt}=H\frac{d}{d\ln a}$, we derive after
some algebra, that
\be
\label{fzz222}
a\frac{dF}{da}+F^{2}+X(a)F
= \frac{3}{2}\Omega_{m}(a)Q(a) \;,
\ee
with
\be
\label{xxa}
X(a)=\frac{1}{2}-\frac{3}{2}w(a)
\left[ 1-\Omega_{m}(a)\right]
\ee
where in order to evaluate the final form of Eq.(\ref{xxa})
we have used Eq.(\ref{domm}).

Now, we consider that the growth index varies with cosmic time.
Transforming equation (\ref{fzz222})
from $a$ to redshift [$\frac{d}{da}=-(1+z)^{-2}\frac{d}{dz}$]
and utilizing Eqs.(\ref{fzz221}), (\ref{domm})
we simply derive the evolution equation
of the growth index $\gamma=\gamma(z)$ (see also~\cite{Pol}).
Indeed this is given by:
\bear
\label{Poll}
&& -(1+z)\gamma^{\prime}{\rm ln}(\Omega_{m})+\Omega_{m}^{\gamma}+
3w(1-\Omega_{m})(\gamma-\frac{1}{2})+\frac{1}{2}\; \nonumber \\
&& =\frac{3}{2}Q\Omega_{m}^{1-\gamma} \;,
\eear
where prime denotes derivative with respect to redshift.
At the present time the above equation becomes
\bear
\label{Poll1}
&& -\gamma^{\prime}(0){\rm ln}(\Omega_{m0})+\Omega_{m0}^{\gamma(0)}+
3w_{0}(1-\Omega_{m0})[\gamma(0)-\frac{1}{2}]+\frac{1}{2}\; \nonumber \\
&&=\frac{3}{2}Q_{0}\Omega_{m0}^{1-\gamma(0)}\;,
\eear
where $Q_{0}=Q(z=0,\kappa)$ and $w_{0}=w(z=0)$.

In this work we phenomenologically parametrize $\gamma(z)$
by the following general relation (see \cite{Por})
\be
\gamma(z)=\gamma_{0}+\gamma_{1}y(z)\;. \label{yz}
\ee
Obviously, the above equation can be viewed as a first order Taylor expansion
around some cosmological quantity such as $a(z)$, $z$ and $\Omega_{m}(z)$.
We would like to stress that for those $y(z)$ functions which satisfy $y(0)=0$
[or $\gamma(0)=\gamma_{0}$] one can write the
parameter $\gamma_{1}$ in terms of $\gamma_{0}$.
In this case [$\gamma^{\prime}(0)=\gamma_{1}y^{\prime}(0)$], using
Eq.(\ref{Poll1}) we obtain
\be
\label{Poll2}
\gamma_{1}=\frac{\Omega_{m0}^{\gamma_{0}}+3w_{0}(\gamma_{0}-\frac{1}{2})
(1-\Omega_{m0})-\frac{3}{2}Q_{0}\Omega_{m0}^{1-\gamma_{0}}+\frac{1}{2}  }
{y^{\prime}(0)\ln  \Omega_{m0}}\;.
\ee

Let us now briefly present various forms of $\gamma(z)$, $\forall z$.
\begin{itemize}

\item Constant growth index (hereafter $\Gamma_{0}$ model): Here
we set $\gamma_{1}$ strictly equal to zero, thus $\gamma=\gamma_{0}$.

\item Expansion around $z=0$ (see \cite{Pol}; hereafter $\Gamma_{1}$ model):
In this case we have $y(z)=z$. Note however, that this parametrization is valid
at relatively low redshifts $0\le z \le 0.5$.
In the statistical
analysis presented below we utilize a constant
growth index, namely $\gamma=\gamma_{0}+0.5\gamma_{1}$ for $z>0.5$.

\item Expansion around $a=1$ (\cite{Bel12,DP11,Ishak09}; hereafter
$\Gamma_{2}$ model): Here
the function $y$ becomes $y(z)=1-a(z)=\frac{z}{1+z}$. Obviously,
at large redshifts $z\gg 1$ we get
$\gamma_{\infty}\simeq \gamma_{0}+\gamma_{1}$.
\end{itemize}
For the $\Gamma_{1}$ and $\Gamma_{2}$
parametrizations one can easily show that $y(0)=0$ and
$y^{\prime}(0)=1$, respectively. As an example, for the case of the $\Lambda$CDM
cosmology with $\gamma_{0}\simeq 6/11$ and
$\Omega_{m0}=0.273$, Eq.(\ref{Poll2}) provides $\gamma_{1}\simeq -0.0478$.
In addition, based on the Starobinsky's $f(R)$ model
with $(\Omega_{m0},\gamma_{0})=(0.273,0.415)$
Gannouji et al. \cite{Gann09} found $\gamma_{1}=-0.21$.


Finally, we should note that the growth index is clearly model dependent via $\gamma_{1}$
as it can be seen by Eqs.(\ref{yz}) and (\ref{Poll2}).
However, Gannouji et al. Ref. \cite{Gann09} found that in
the case of the Starobisky $f(R)$ model
the corresponding growth rate of clustering $f(z)$ is not really affected
by the scale especially up to $z=2$ (see their Fig. 2).
In addition, we have demonstrated that the allowed deviation from $\Lambda$CDM is relatively
small in the cases
considered ($b$ less than 0.5 at 2$\sigma$ in most cases) and therefore any
allowed scale dependence
of the growth is minor. This implies that the use of the value of the
measured $f\sigma_8$ can be used without sacrifice of accuracy.

Another issue concerning nonlinear effects is that in this work
we utilize $\kappa=1/\lambda=0.1h$Mpc$^{-1}$ which
corresponds to $\lambda=10h^{-1}$Mpc. Note that
the power-spectrum normalization $\sigma_{8}$ which is
the rms mass fluctuation on $R_{8}=8 h^{-1}$ Mpc corresponds to
$\kappa =0.125h$Mpc$^{-1}$. On the other hand it has
been common practice to assume that the
shape of the power spectrum recovered from galaxy surveys
matches the linear matter power spectrum shape on
scales $\kappa \le 0.15h$Mpc$^{-1}$ (\cite{Smith:2002dz},
\cite{Tegmark:2003uf}, see also the discussion in section 4 of
\cite{Percival:2006gt}). Obviously the choice of
$\kappa=0.1h$Mpc$^{-1}$ insures that we are treating the liner regime.
Of course we have repeated our analysis for different values of $\kappa$
and we confirm Gannouji et al. results, ie that
small variations around $\kappa=0.1h$Mpc$^{-1}$ do not really affect
the qualitative evolution of the growth rate of clustering and thus
of $\gamma$. Furthermore, we found that the evolution of $G_{eff}(z)$ is almost completely unaffected for different
values of $\kappa$, see for example Fig.~\ref{fig:geff} (top) and Fig.~\ref{fig:geff001}. 

Nevertheless, we do anticipate a minor contribution of non-linear effects even on these scales at a level
less than a few percent \cite{Angulo:2007fw}.
These effects would tend to slightly amplify
the value of $\gamma$ and increase the error bars correspondingly by less than a few percent.

\begin{table*}[t!]
\caption[]{Summary of the observed growth rate and references.}
\tabcolsep 6pt
\begin{tabular}{cccc}
\hline
Index & $z$ & growth rate $(\fs_{obs})$& Refs.\\ \hline \hline
1&0.02 & $0.360\pm 0.040$& \cite{Hud12}\\
2&0.067 & $0.423\pm 0.055$& \cite{Beutler}\\
3&0.17 & $0.510\pm 0.060$& \cite{Perc04,Song09}\\
4&0.35 & $0.440\pm 0.050$& \cite{Song09, Teg06} \\
5&0.77 & $0.490\pm 0.180$& \cite{Song09, Guzzo08}\\
6&0.25 & $0.351\pm 0.058$&\cite{Sam11}\\
7&0.37 & $0.460\pm 0.038$& \cite{Sam11}\\
8&0.22 & $0.420\pm 0.070$&\cite{Blake}\\
9&0.41 & $0.450\pm 0.040$& \cite{Blake}\\
10&0.60 & $0.430\pm 0.040$& \cite{Blake}\\
11&0.78 & $0.380\pm 0.040$& \cite{Blake}\\
12&0.57 & $0.427\pm 0.066$& \cite{Reid12}\\
13&0.30 & $0.407\pm 0.055$& \cite{Tojeiro:2012rp}\\
14&0.40 & $0.419\pm 0.041$& \cite{Tojeiro:2012rp}\\
15&0.50 & $0.427\pm 0.043$& \cite{Tojeiro:2012rp}\\
16&0.60 & $0.433\pm 0.067$& \cite{Tojeiro:2012rp}\\
\end{tabular}
\end{table*}

\section{Observational constraints}
In the following we briefly present some details of the
statistical method and on the observational sample
that we adopt in order to constrain the free parameters of the growth index,
presented in the previous section.

\subsection{The Growth data}
The growth data that we utilize in this article based on
the PSCz, 2dF, VVDS, SDSS, 6dF, 2MASS, BOSS and {\em WiggleZ} galaxy surveys,
for which their combination parameter of the growth rate of structure,
$F(z)$, and the redshift-dependent rms fluctuations of the linear
density field, $\sigma_8(z)$,
is available as a function of redshift, $F(z)\sigma_{8}(z)$.
The $F\sigma_{8}\equiv \fs$ estimator is almost a model-independent
way of expressing the observed growth history of the universe (see \cite{Song09}).
Indeed the observed growth rate of structure ($F_{obs}=\beta {\cal B}$) is derived
from the redshift space distortion parameter $\beta(z)$ and the linear bias ${\cal B}$.
Observationally, using the anisotropy of the correlation function one can estimate the $\beta(z)$ parameter. On the other hand, the linear bias factor can be defined as the ratio of the variances of the tracer (galaxies, QSOs etc) and underlying mass density fields, smoothed at $8h^{-1}$ Mpc ${\cal B}(z)=\sigma_{8,tr}(z)/\sigma_{8}(z)$, where $\sigma_{8,tr}(z)$ is measured directly from the sample. Combining the above definitions we arrive at $\fs\equiv F \sigma_{8}=\beta \sigma_{8,tr}$. We would like to point that the different cosmologies (including those of modified gravity)
enter only weakly in the observational determination of $\beta(z)$ (and thus of $\fs$), through the definition of distances. In Table I we quote the precise numerical values of the data points with the corresponding errors and references.

\subsection{The overall Likelihood analysis}
In order to constrain the cosmological parameters and
the growth index of the $f(R)$ models one needs to
perform a joint likelihood analysis, involving the
cosmic expansion data such as SnIa and BAO, CMB shift parameter
together with the growth data. Up to now, due to
the large errors of the growth data with respect
to the cosmic expansion data, various authors preferred to
constrain first $\Omega_{m0}$ using SnIa/BAO/CMB
and then to use the growth data alone.
Of course, armed with the recent high quality growth data
it would be worthwhile to simultaneously
constrain $(\Omega_{m0},b,\gamma)$.
In particular, we use the {\em Union 2.1} set of 580 SnIa of
Suzuki et al. \cite{Suzuki:2011hu}
\footnote{The SnIa data can be found in
http://supernova.lbl.gov/Union/ and
in \cite{Suzuki:2011hu}},
and the observed BAOs. For simplicity, but without loss of generality, we only considered the case where the covariance matrix of the SnIa data is diagonal. The BAO data are given in terms of the parameter $d_z(z)=\frac{l_{BAO}(z_{drag})}{D_V(z)}$, where
$l_{BAO}(z_{drag})$ is the BAO scale at the drag redshift, assumed
known from CMB measurements, and \cite{Perc10}
\be
D_V(z)=\left[(1+z)^2 D_A(z)^2 \frac{c\,z}{H(z)}\right]^{1/3}
\ee
is the usual volume distance.

\begin{table}
\begin{center}
\caption{The BAO data used in this analysis. The first six data points are volume averaged and correspond to Table 3 of \cite{Blake:2011en}. Their inverse covariance Matrix is given by (\ref{covbao}).  \label{table2}}
\begin{tabular}{| c | c | cc | ccc |}
\multicolumn{1}{c}{} & \multicolumn{1}{c}{6dF} & \multicolumn{2}{c}{SDSS}
 & \multicolumn{3}{c}{WiggleZ}  \\\hline
$z$ & 0.106 & 0.2 & 0.35 & 0.44 & 0.6 & 0.73  \\\hline
$d_z$ & 0.336 & 0.1905 & 0.1097 & 0.0916 & 0.0726 & 0.0592  \\\hline
$\Delta d_z$ & 0.015 & 0.0061 & 0.0036 & 0.0071 & 0.0034 & 0.0032 \\\hline
 \end{tabular}
\end{center}
\end{table}

In this analysis we use the 6dF, the SDSS and WiggleZ BAO data shown in Table \ref{table2}. The WiggleZ collaboration \cite{Blake:2011en} has measured the baryon acoustic scale at three different redshifts, complementing previous data at lower redshift obtained by SDSS and 6DFGS \cite{Perc10}.

The chi-square is given by
\begin{equation} \label{chi2bao}
\chi^2_{BAO}= \sum_{i,j} [d_i - d(z_i)]C^{-1}_{ij}[d_j-d(z_j)],
\end{equation}
where the indices $i,j$ are in growing order in $z$, as in Table \ref{table2}.
For the first six points, $C_{ij}^{-1}$ was obtained from the covariance data in \cite{Blake:2011en} in terms of $d_z$:
\begin{equation}\label{covbao}
 C_{ij}^{-1}= \left(
\begin{array}{cccccc}
 4444 & 0. & 0. & 0. & 0. & 0. \\
 0. & 30318 & -17312 & 0. & 0. & 0. \\
 0. & -17312 & 87046 & 0. & 0. & 0. \\
 0. & 0. & 0. & 23857 & -22747 & 10586 \\
 0. & 0. & 0. & -22747 & 128729 & -59907 \\
 0. & 0. & 0. & 10586 & -59907 & 125536
\end{array}
\right)
\end{equation}

The positions of CMB acoustic peaks are affected by the expansion
history of the Universe from the decoupling epoch to today.
In order to quantify the shift of acoustic peaks we use the
data points $(l_a, {\cal R}, z_{cmb})$ of Ref.~\cite{Hinshaw:2012fq} (WMAP9),
where $l_a$ and ${\cal R}$ are two CMB shift
parameters and $z_{cmb}$ is the redshift at decoupling.

For the redshift $z_{cmb}$ there is a fitting formula by
Hu and Sugiyama \cite{Hu:1995en}:
\begin{equation}
z_{cmb}=1048\left(1+0.00124\omega_{b}^{-0.738}\right)
\left(1+g_{1}\omega_{m}^{g_{2}}\right)\,,\label{zdec}
\end{equation}
where
$g_{1}=0.0783\omega_{b}^{-0.238}/\left(1+
39.5\omega_{b}^{0.763}\right)$,
$g_{2}=0.560/\left(1+21.1\omega_{b}^{1.81}\right)$,
$\omega_b \equiv \Omega_{b0}h^2$, and
$\omega_m \equiv \Omega_{m0}h^2$
($h$ correspond to the uncertainty of the Hubble
parameter $H_0$ today, i.e. $H_0=100\,h$\,km\,sec$^{-1}$\,Mpc$^{-1}$).

For a flat prior, the 9-year WMAP data (WMAP9) measured
best-fit values are \cite{Hinshaw:2012fq}
\begin{eqnarray}
\hspace{-.5cm}
\bm{{\bar V}}_{\rm CMB} &=& \left(\begin{array}{c}
l_a \\
{\cal R}\\
{z_{cmb}}\end{array}
\right)=
\left(\begin{array}{c}
302.40 \\
1.7246 \\
1090.88  \end{array}
\right)\,.
\label{cmbdat}
\end{eqnarray}
The corresponding inverse covariance matrix
is \cite{Hinshaw:2012fq}
\begin{eqnarray}
\bm{C}_{\rm CMB}^{-1}=\left(
\begin{array}{ccc}
3.182 & 18.253  & -1.429  \\
18.253& 11887.879& -193.808\\
-1.429& -193.808& 4.556
\end{array}
\right)\,.
\end{eqnarray}
We thus define
\begin{eqnarray}
\bm{X}_{\rm CMB} &=& \left(\begin{array}{c}
l_a-302.40\\
{\cal R} - 1.7246 \\
z_{cmb}-1090.88\end{array}
\right)\,,
\end{eqnarray}
and construct the contribution of CMB to $\chi^2$ as
\be
\chi^2_{\rm CMB}=\bm{X}_{\rm CMB}^{T}\,
\bm{C}_{\rm CMB}^{-1}\,\bm{X}_{\rm CMB}\,.
\ee

Notice that $\chi^2_{\rm CMB}$ depends on the parameters
($\Omega_{m0}$, $\Omega_{b0}$, $h$).
The density parameter of radiation today is
\be
\Omega_{r0}=\Omega_{\gamma0}
(1+0.2271 N_{\rm eff})\,,
\label{Omer}
\ee
where $\Omega_{\gamma0}$ is the photon density parameter
and $N_{\rm eff}$ is the relativistic degrees of freedom.
We adopt the standard values
$\Omega_{\gamma0}=2.469 \times 10^{-5}\,h^{-2}$
and $N_{\rm eff}=3.04$ \cite{Hinshaw:2012fq}.
Concerning the constraints on the parameters assuming
a fixed $H_{0}$ value we should note that the statistical analysis
does not depend on an a priory selected value of $H_{0}$.
First, the SnIa distance
moduli are always normalized using the internally determined $H_{0}$.
On the other hand one of the merits of using the shift
parameter in cosmological studies is
that its dependence on the Hubble constant is almost negligible (for details
see \cite{MelNes}). Indeed as it can be shown from
Eq.(\ref{zdec} we use the normalized cosmological parameters
$\omega_{m}=\Omega_{m0}h^2$ and $\omega_{b}=\Omega_{b0}h^2$.
In this context, the $H_{0}$-dependence does it enter in the
analysis of the CMB shift parameter via $\Omega_{r0}$,
but small variations around $\sim 70$ km/sec/Mpc are not expected to
affect the qualitative results.

\begin{figure*}[t!]
\centering
\vspace{0cm}\rotatebox{0}{\vspace{0cm}\hspace{0cm}\resizebox{0.3\textwidth}{!}{\includegraphics{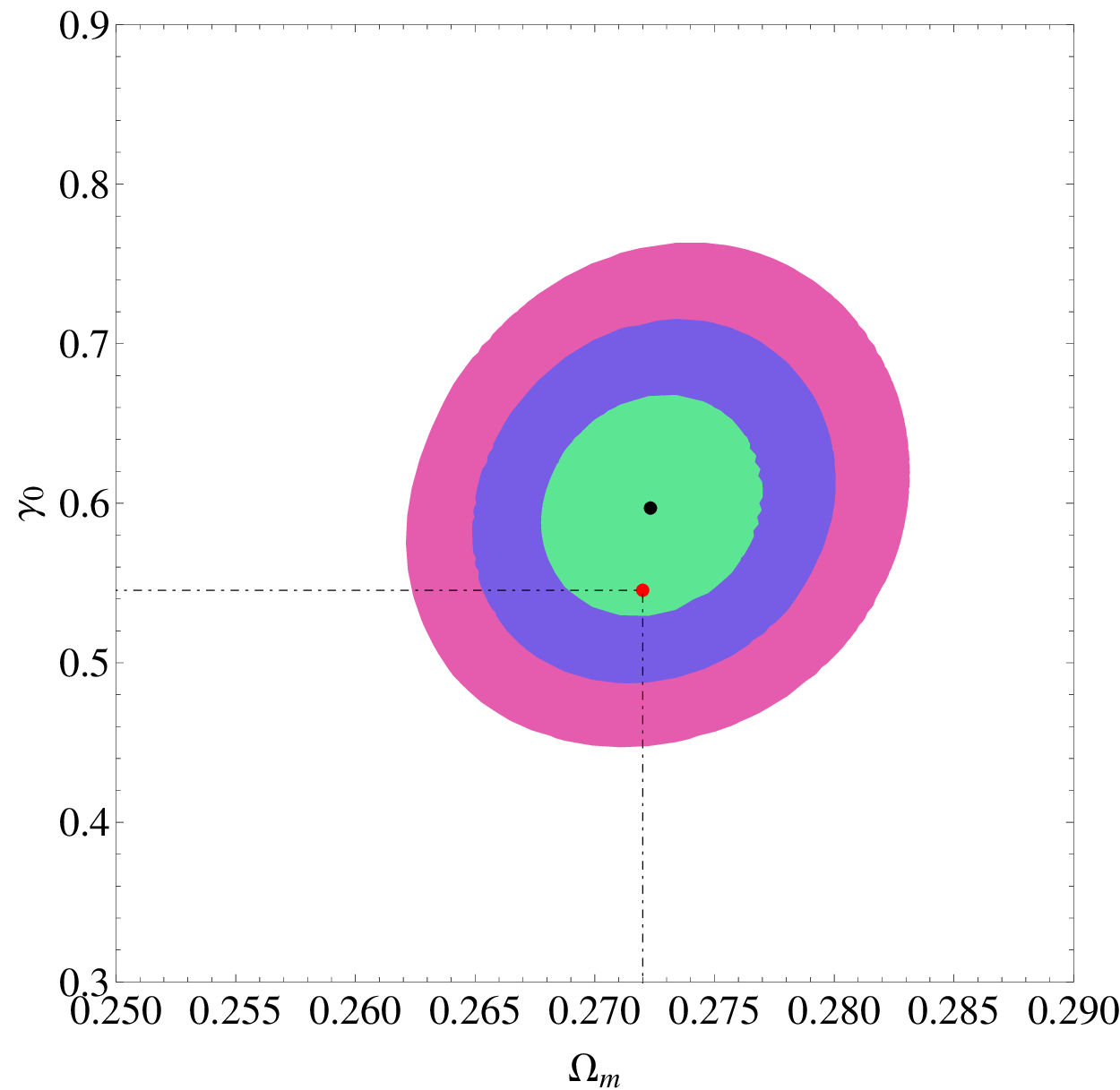}}}
\vspace{0cm}\rotatebox{0}{\vspace{0cm}\hspace{0cm}\resizebox{0.3\textwidth}{!}{\includegraphics{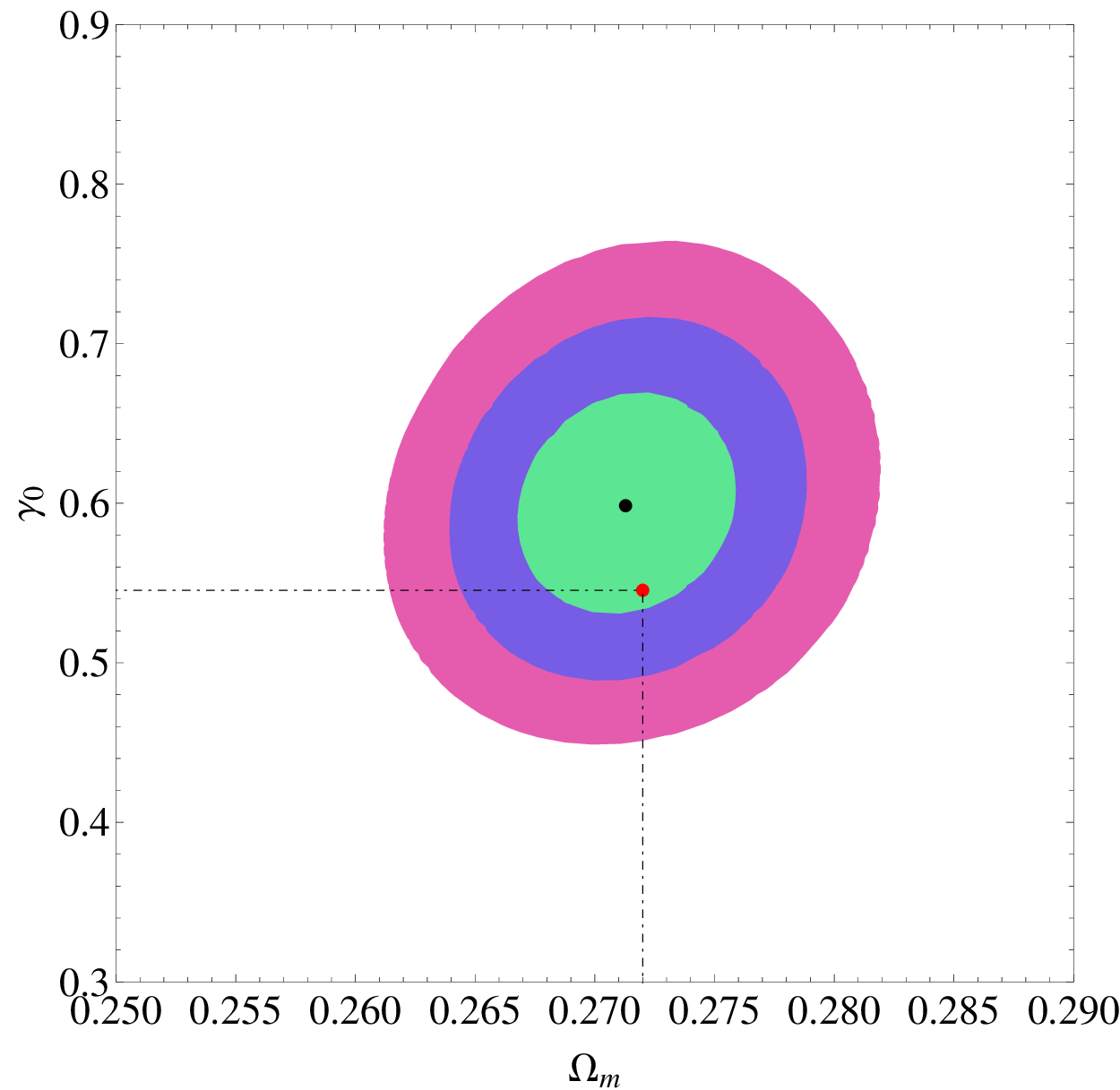}}}
\vspace{0cm}\rotatebox{0}{\vspace{0cm}\hspace{0cm}\resizebox{0.3\textwidth}{!}{\includegraphics{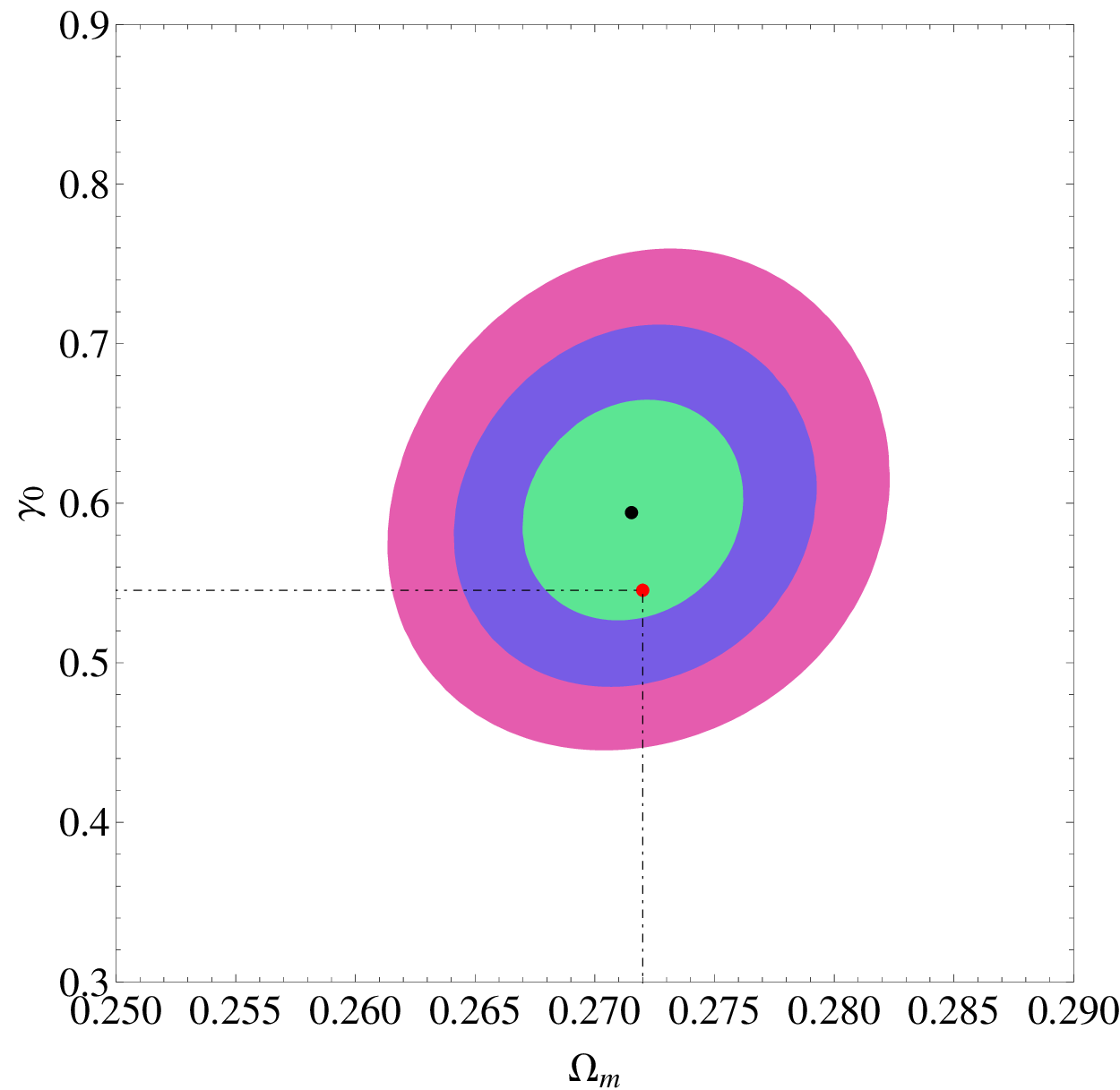}}}
\caption{{\em Left:} Likelihood contours for $\chi^2$ equal to 2.30, 6.18 and 11.83, corresponding to 1$\sigma$, 2$\sigma$ and $3\sigma$ confidence levels, in the
$(\Omega_{m0},\gamma)$ plane using a $\Lambda$CDM expansion model.
{\em Middle and Right:} Here we show the corresponding
contours in the case of $f(R)$ models ($f_{1}$CDM -  middle panel and $f_{2}$CDM -right panel). In all cases the red point corresponds to $(\Omega_{m0},\gamma)=(0.272,6/11)$. In this plot and in the ones that follow we have set the parameters that are not shown (eg $b$) to their best fit values for the corresponding model (see Table \ref{tab:growth1}).Here we use $n=1$.\label{figcon1}}
\end{figure*}

\begin{figure}[t!]
\centering
\vspace{0cm}\rotatebox{0}{\vspace{0cm}\hspace{0cm}\resizebox{0.5\textwidth}{!}{\includegraphics{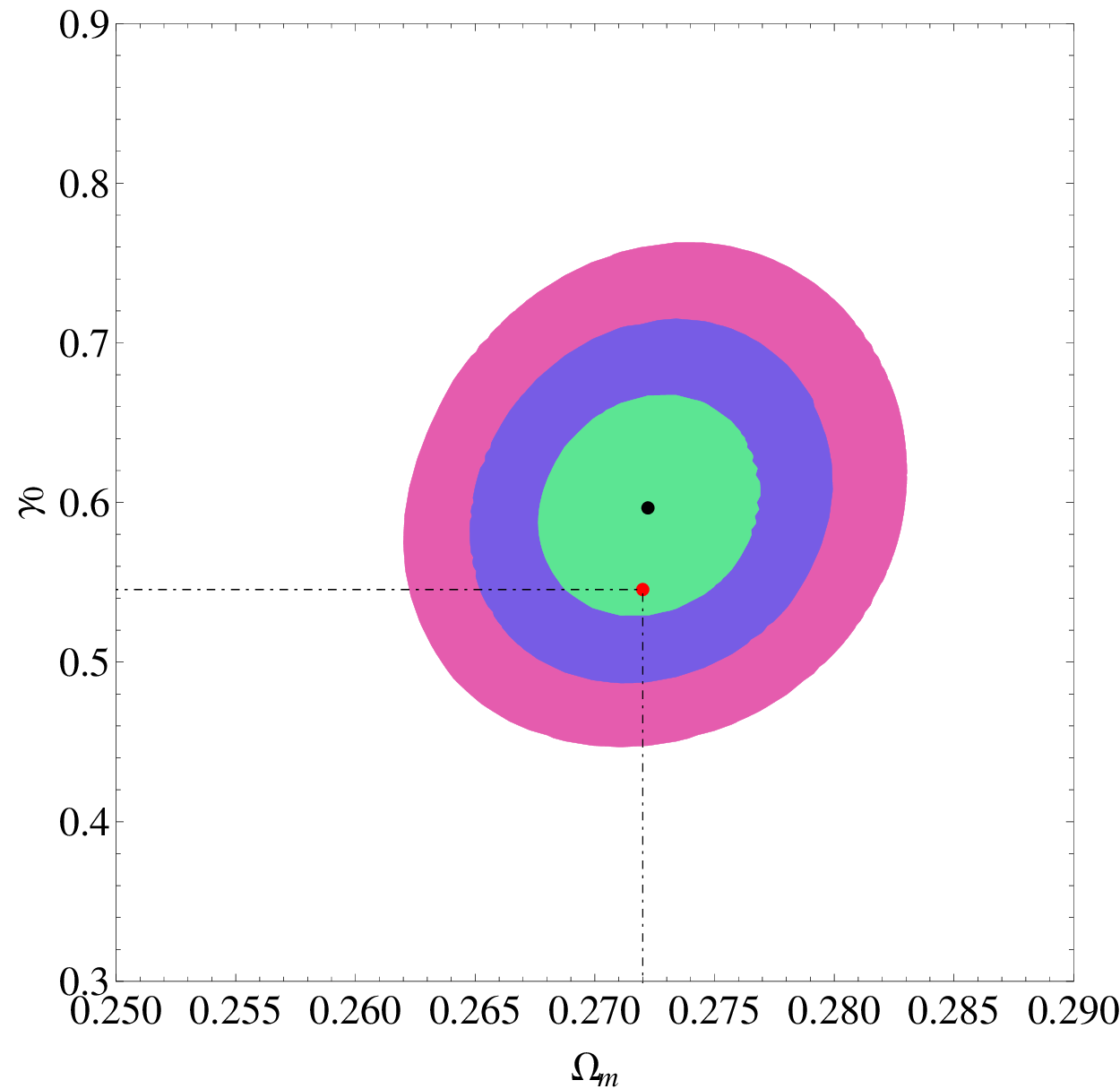}}}
\caption{The likelihood contours for $\chi^2$ equal to 2.30, 6.18 and 11.83, corresponding to 1$\sigma$, 2$\sigma$ and $3\sigma$ confidence levels, in the
$(\Omega_{m0},\gamma)$ plane in the case of the $f_{1}$CDM model for $n=2$. In all cases the red point corresponds to $(\Omega_{m0},\gamma)=(0.272,6/11)$. Clearly, our results remain mostly unaffected by the choice of a particular $n$. \label{figcon1n2}}
\end{figure}

\begin{figure*}[t!]
\centering
\vspace{0cm}\rotatebox{0}{\vspace{0cm}\hspace{0cm}\resizebox{0.75\textwidth}{!}{\includegraphics{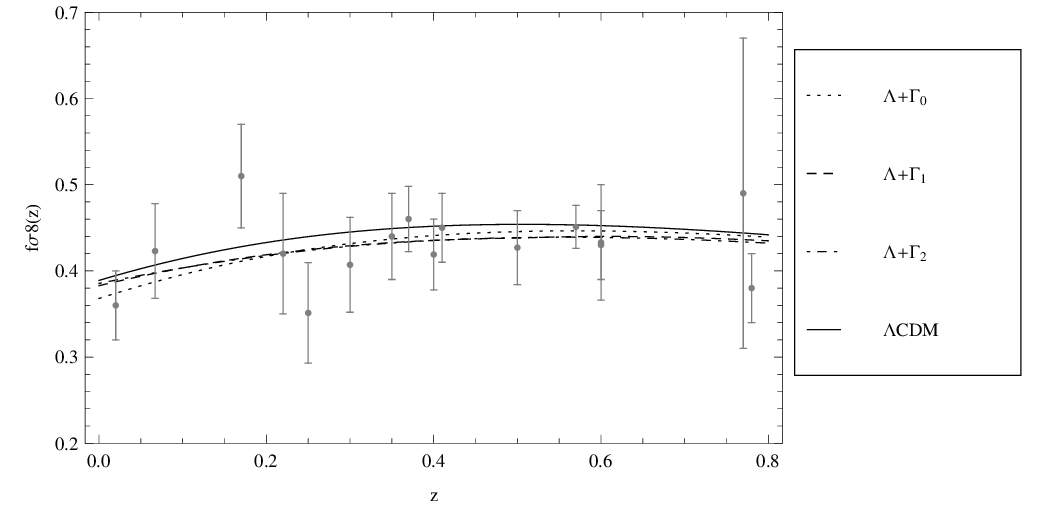}}}
\vspace{0cm}\rotatebox{0}{\vspace{0cm}\hspace{0cm}\resizebox{0.75\textwidth}{!}{\includegraphics{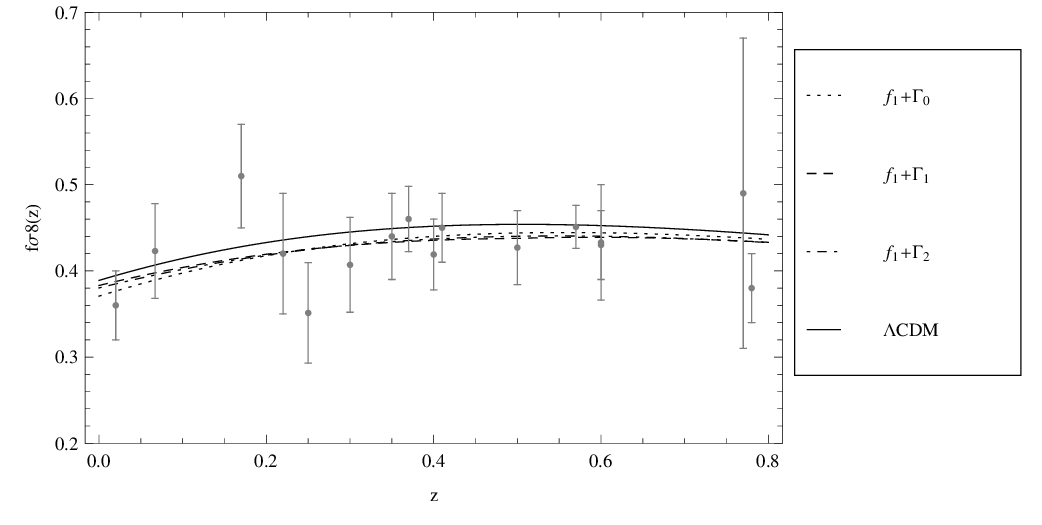}}}
\vspace{0cm}\rotatebox{0}{\vspace{0cm}\hspace{0cm}\resizebox{0.75\textwidth}{!}{\includegraphics{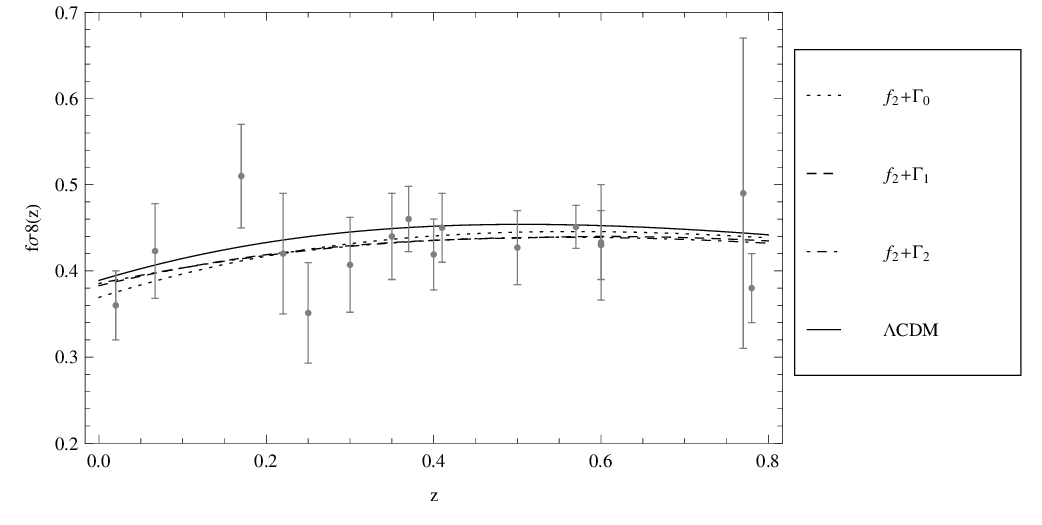}}}
\caption{Comparison of the observed and theoretical evolution of the growth rate $\fs(z)=F(z)\sigma_{8}(z)$. Top: The dotted, dashed and dot-dashed lines correspond to the best-fit $\Gamma_0$, $\Gamma_1$ and $\Gamma_2$ parametrizations with the background expansion given by $\Lambda$CDM, while the black line to the exact solution of Eq.~(\ref{odedelta}) for $\fs(z)$ for the $\Lambda$CDM model for $\Omega_{m0}=0.273$
\cite{Hinshaw:2012fq}. Middle and Bottom:
The dotted, dashed and dot-dashed lines correspond to the best-fit $\Gamma_0$, $\Gamma_1$ and $\Gamma_2$ parametrizations with the background expansion given by $f_{1}$CDM and $f_{2}$CDM respectively, while the black line to the exact solution of Eq.~(\ref{odedelta}) for $\fs(z)$ for the $\Lambda$CDM model for $\Omega_{m0}=0.272$ \cite{Hinshaw:2012fq}. In all cases we utilize
$\sigma_{8}=0.8$ and $n=1$. \label{growthrate}}
\end{figure*}

The overall likelihood function\footnote{Likelihoods are
normalized to their maximum values. In the present analysis we
always report $1\sigma$ uncertainties on the fitted parameters.}
is given by the product of the individual
likelihoods according to:
\begin{equation}\label{eq:overalllikelihood}
{\cal L}_{tot}({\bf p_{1}},{\bf p_{2}})={\cal L}_{E} ({\bf p_{1}})
\times {\cal L}_{f}({\bf p_{1}},{\bf p_{2}})
\end{equation}
where ${\cal L}_{f}$ refers to the dynamical probe likelihood fit and
\begin{equation}\label{eq:overalllikelihood1}
{\cal L}_{E} ({\bf p_{1}})=
{\cal L}_{SNIa}\times {\cal L}_{BAO} \times {\cal L}_{CMB} \;.
\end{equation}
The vectors ${\bf p_{1}}$, ${\bf p_{2}}$
contain the free parameters of the $f(R)$ model
and depend on the model. In particular,
the essential free parameters that enter in the
theoretical expectation are:
${\bf p_{1}} \equiv (\Omega_{m0},b)$ and
${\bf p_{2}} \equiv (\gamma_{0},\gamma_{1})$.
Note that in the case of the $\Lambda$CDM we have ${\bf p_{1}} \equiv \Omega_{m0}$. Also, in all cases we have set $\sigma_{8}=0.8$.

Since likelihoods are defined as ${\cal L}\propto
\exp{\left(-\chi^2/2\right)}$, this translates into an addition for the
joint $\chi^2_{tot}$ function:
\begin{equation}\label{eq:overalllikelihoo}
\chi^{2}_{tot}({\bf p_{1}},{\bf p_{2}})=\chi^{2}_{E}({\bf p_{1}})+
\chi^{2}_{f}({\bf p_{1}},{\bf p_{2}})
\end{equation}
with
\begin{equation}\label{eq:overalllikelihoo1}
\chi^{2}_{E}({\bf p_{1}})=
\chi^{2}_{SNIa}+\chi^{2}_{BAO}+\chi^{2}_{CMB} \;.
\end{equation}
The minimization of the $\chi^2_{tot}$ was
done in Mathematica\texttrademark. Note that the $\chi^{2}_{f}$ is given by
\be
\label{Likel}
\chi^{2}_{f}({\bf p_{1}},{\bf p_{2}},z_{i})=
\sum_{i=1}^{N_{f}} \left[ \frac{\fs_{obs}(z_{i})-
\fs({\bf p_{1}},{\bf p_{2}},z_{i})}
{\sigma_{i}}\right]^{2}
\ee
where $\sigma_{i}$ is the observed growth rate uncertainty.

To this end since $N/n_{fit}>40$ we will use, the
relevant to our case,
{\em corrected} Akaike information criterion \cite{Akaike1974}, defined,
for the case of Gaussian errors, as:
\be
{\rm AIC}=\chi^2_{min}+2n_{fit}
\ee
where $N=N_{EXP}+N_{f}$, $n_{fit}$ is the number of free parameters. A smaller value of AIC indicates a better model-data fit. However, small differences in AIC are not necessarily significant and therefore, in order to assess, the effectiveness of the different models in reproducing the data, one has to investigate the model pair difference $\Delta$AIC$ = {\rm AIC}_{y} - {\rm AIC}_{x}$. The higher the value of $|\Delta{\rm AIC}|$, the higher the evidence against the model with higher value of ${\rm AIC}$, with a difference $|\Delta$AIC$| \magcir 2$ indicating a positive such evidence and $|\Delta$AIC$| \magcir 6$ indicating a strong such evidence, while a value $\mincir 2$ indicates consistency among the two comparison models. A numerical summary of the statistical analysis for the  background expansion models as well as for the various $\gamma(z)$ parametrizations is shown in Table \ref{tab:growth1}.

At this point we should stress that in this paper we
only use the shift parameter and don't use the full CMB likelihood.
The reason for this is that our analysis has demonstrated self-consistently
that only small deviations from $\Lambda$CDM are allowed. Thus
the use of the shift parameter for this range of small deviations
is expected to be an acceptable approximation to the more accurate
(but also more complicated) full CMB likelihood approach.

Furthermore, we have decided to utilize (as many authors
did in the past)
the CMB shift parameter which is a valid and frequently used
tool in this kind of studies, especially over the last decade.
The robustness of the shift parameter was tested and
discussed in Refs\cite{ElgCor} and it has been found that
the shift parameter changes when massive neutrinos are included
(which is not our case here)
or when there is a strongly varying equation of state parameter
(the $f(R)$ models remain close to $\Lambda$CDM).

\subsubsection{Constant growth index}
First of all we utilize the $\Gamma_{0}$ parametrization
($\gamma=\gamma_{0}$, $\gamma_{1}=0$: see section 3A).
Therefore, the corresponding $f_1$CDM and $f_2$CDM statistical vectors ${\bf p}_{2}$ contain only three free parameters namely, ${\bf p_{2}} \equiv ({\bf p_{1}},\gamma_{0},0)$ where ${\bf p_{1}} \equiv (\Omega_{m0},b)$. Accordingly, if
we consider the $\Lambda$CDM model then ${\bf p_{1}} \equiv \Omega_{m0}$,
implying that the vector ${\bf p_{2}}$ includes two free parameters.

Our main results are listed in Table \ref{tab:growth1}, where we quote the best fit parameters with the corresponding 1$\sigma$ uncertainties, for three different expansion models. In Figure \ref{figcon1} we present the 1$\sigma$, 2$\sigma$ and $3\sigma$ confidence levels in the $(\Omega_{m0},\gamma)$ plane. It becomes evident that using the most recent growth data-set together with the expansion cosmological data we can place strong constraints on $(\Omega_{m0},\gamma)$.
In all cases the best fit value
$\Omega_{m0}=0.272 \pm 0.003$ is in a very good agreement with that provided
by WMAP9+SPT+ACT ($\Omega_{m0}=0.272$; Hinshaw et al.  \cite{Hinshaw:2012fq}).

\begin{table*}[ht]
\caption[]{Statistical results for the combined growth data (see Table I): The $1^{st}$ column indicates the expansion model, the $2^{nd}$ column corresponds to
$\gamma(z)$ parametrizations appearing in section 3A. $3^{rd}$ and $4^{rth}$ columns provide the $\Omega_{m0}$ and $b$ best
values. The $5^{th}$ and $6^{th}$ columns show the $\gamma_{0}$ and
$\gamma_1$ best fit values. In all cases we used $\sigma_{8}=0.8$. The remaining columns present the goodness-of-fit statistics
($\chi^{2}_{min}$, AIC and $|\Delta$AIC$|=|{\rm AIC}_{\Lambda}-{\rm AIC}_{f(R)}|$).}\tabcolsep 4.5pt
\vspace{1mm}
\begin{tabular}{ccccccccc} \hline \hline
Exp. Model & Param. Model & $\Omega_{m0}$ & $b$ & $\gamma_{0}$& $\gamma_{1}$& $\chi_{min}^{2}$ &${\rm AIC}$& $|\Delta$AIC$|$ \vspace{0.05cm}\\ \hline
$\Lambda$CDM&$\Gamma_{0}$& $0.272\pm0.003$ &  & $0.597\pm 0.046$& $0$              &574.227& 578.227 & 0   \vspace{0.01cm}\\
            &$\Gamma_{1}$& $0.272\pm0.003$ &  & $0.567\pm 0.066$& $0.116\pm 0.191$ &573.861& 579.861 & 1.634 \vspace{0.01cm}\\
            &$\Gamma_{2}$& $0.272\pm0.003$ &  & $0.561\pm 0.068$& $0.183\pm 0.269$ &573.767& 579.767 & 1.540 \vspace{0.15cm}\\
$f_{1}$CDM-\cite{Hu07} &$\Gamma_{0}$& $0.271\pm0.003$ & $0.111\pm0.140$& $0.598\pm 0.046$& 0                &573.855& 579.855 & 1.628 \vspace{0.01cm}\\
                       &$\Gamma_{1}$& $0.271\pm0.003$ & $0.109\pm0.142$& $0.573\pm 0.068$& $0.097\pm 0.195$ &573.633& 581.633 & 3.406 \vspace{0.01cm}\\
                       &$\Gamma_{2}$& $0.271\pm0.003$ & $0.109\pm0.142$& $0.579\pm 0.070$& $0.101\pm 0.275$ &573.585& 581.585 & 3.358 \vspace{0.15cm}\\
$f_{2}$CDM-\cite{Starobinsky-2007}&$\Gamma_{0}$& $0.272\pm0.005$ & $0.292\pm 0.647$& $0.594\pm0.047$& 0               &$574.178$& 580.178 & 1.951 \vspace{0.01cm}\\
                                  &$\Gamma_{1}$& $0.272\pm0.005$ & $0.150\pm 1.355$& $0.567\pm0.066$& $0.113\pm0.199$ &$573.857$& 581.857 & 3.630 \vspace{0.01cm}\\
                                  &$\Gamma_{2}$& $0.272\pm0.005$ & $0.149\pm 1.261$& $0.561\pm0.068$& $0.179\pm0.279$ &$573.765$& 581.765 & 3.538 \vspace{0.01cm}\\
\hline\hline
\label{tab:growth1}
\end{tabular}
\end{table*}

Concerning the $\Lambda$CDM expansion model (see the right panel
of Fig.\ref{figcon1}) our growth index results are
in agreement within $1\sigma$ errors, to those of
Samushia et al. \cite{Sam11} who found $\gamma=0.584\pm 0.112$ and to those of \cite{Bass} who obtained $\Omega_{m0}=0.273\pm 0.011$ and $\gamma=0.586\pm 0.080$.
However, our best-fit value $\gamma=0.597\pm 0.046$ is
somewhat greater from the theoretically predicted value of $\gamma_{\Lambda} \simeq 6/11$ (see lines in the right panel of Fig. \ref{figcon1}).
Such a small discrepancy between the theoretical $\Lambda$CDM and
observationally fitted value of $\gamma$ has also been
found by other authors. For example, Di Porto \& Amendola \cite{Port08}
obtained $\gamma=0.60^{+0.40}_{-0.30}$, Gong \cite{Gong10} measured $\gamma=0.64^{+0.17}_{-0.15}$ while Nesseris \& Perivolaropoulos \cite{Nes08}, found $\gamma=0.67^{+0.20}_{-0.17}$. Recently, Basilakos \& Pouri \cite{Por} and Hudson \& Turnbull \cite{Hud12} using a similar analysis found $\gamma=0.602\pm 0.05$ and $\gamma=0.619\pm 0.054$ respectively. In this context, Samushia et al. \cite{Samnew12} obtained $\gamma=0.65\pm 0.05$.

\begin{figure*}[t!]
\centering
\vspace{0cm}\rotatebox{0}{\vspace{0cm}\hspace{0cm}\resizebox{0.45\textwidth}{!}{\includegraphics{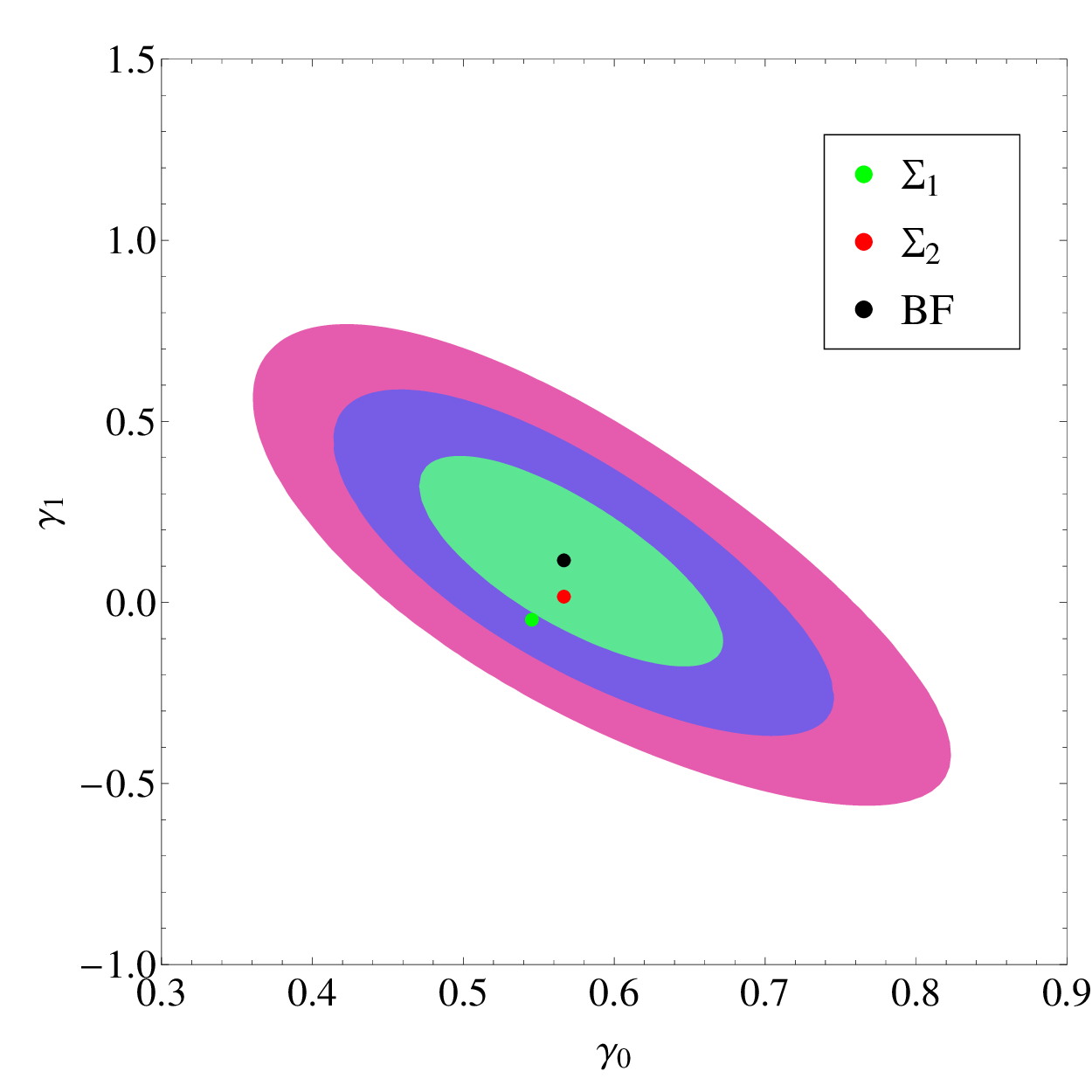}}}
\vspace{0cm}\rotatebox{0}{\vspace{0cm}\hspace{0cm}\resizebox{0.45\textwidth}{!}{\includegraphics{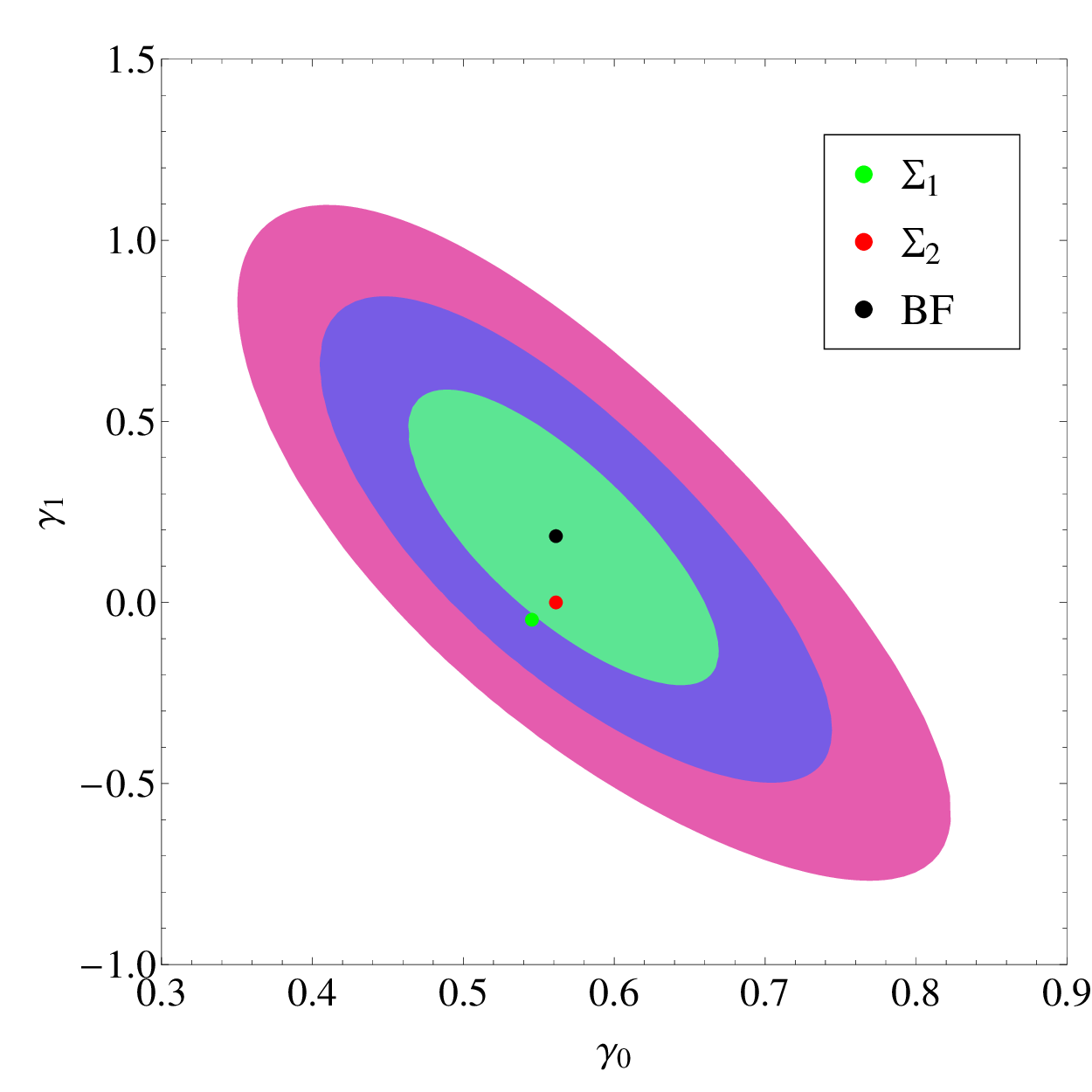}}}
\vspace{0cm}\rotatebox{0}{\vspace{0cm}\hspace{0cm}\resizebox{0.45\textwidth}{!}{\includegraphics{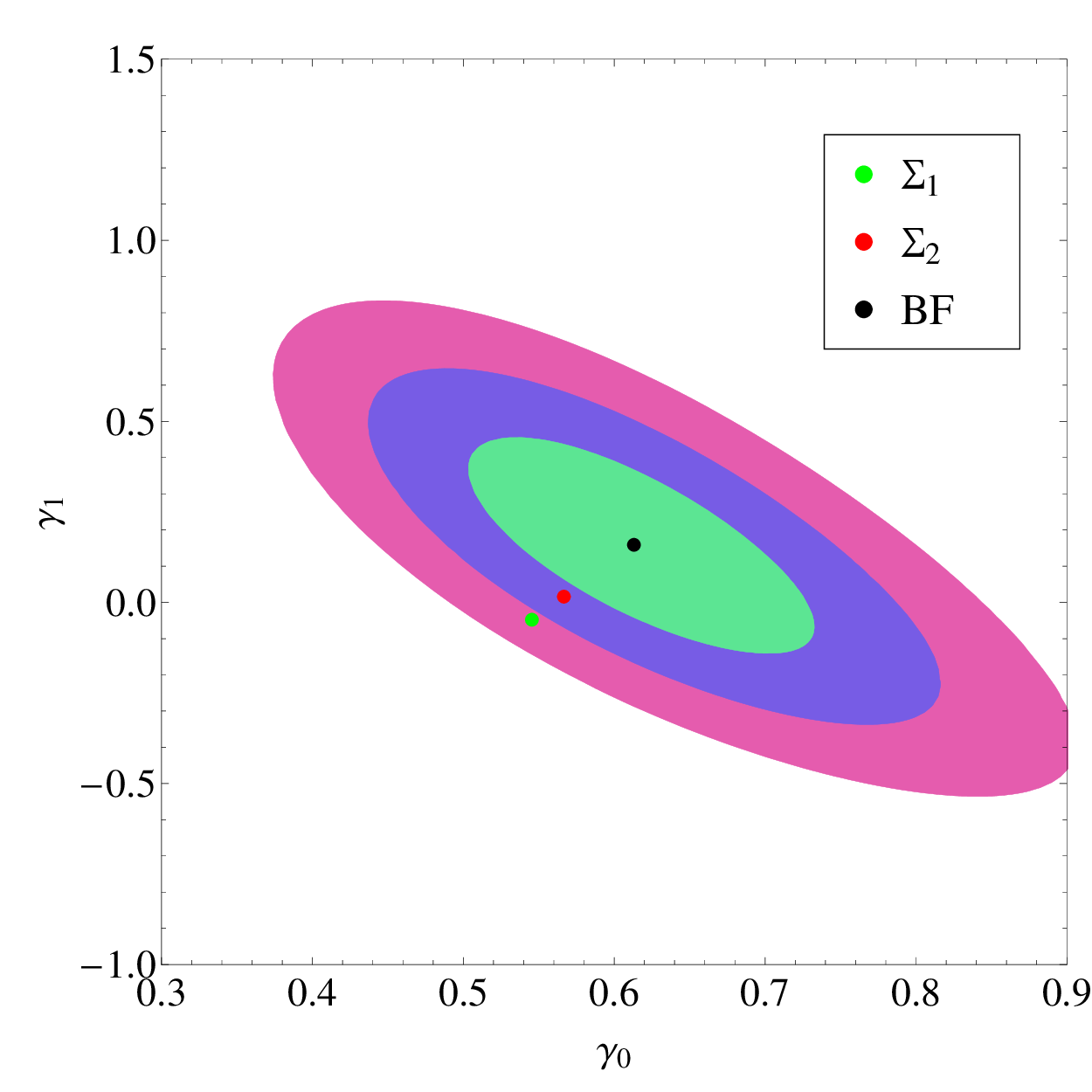}}}
\vspace{0cm}\rotatebox{0}{\vspace{0cm}\hspace{0cm}\resizebox{0.45\textwidth}{!}{\includegraphics{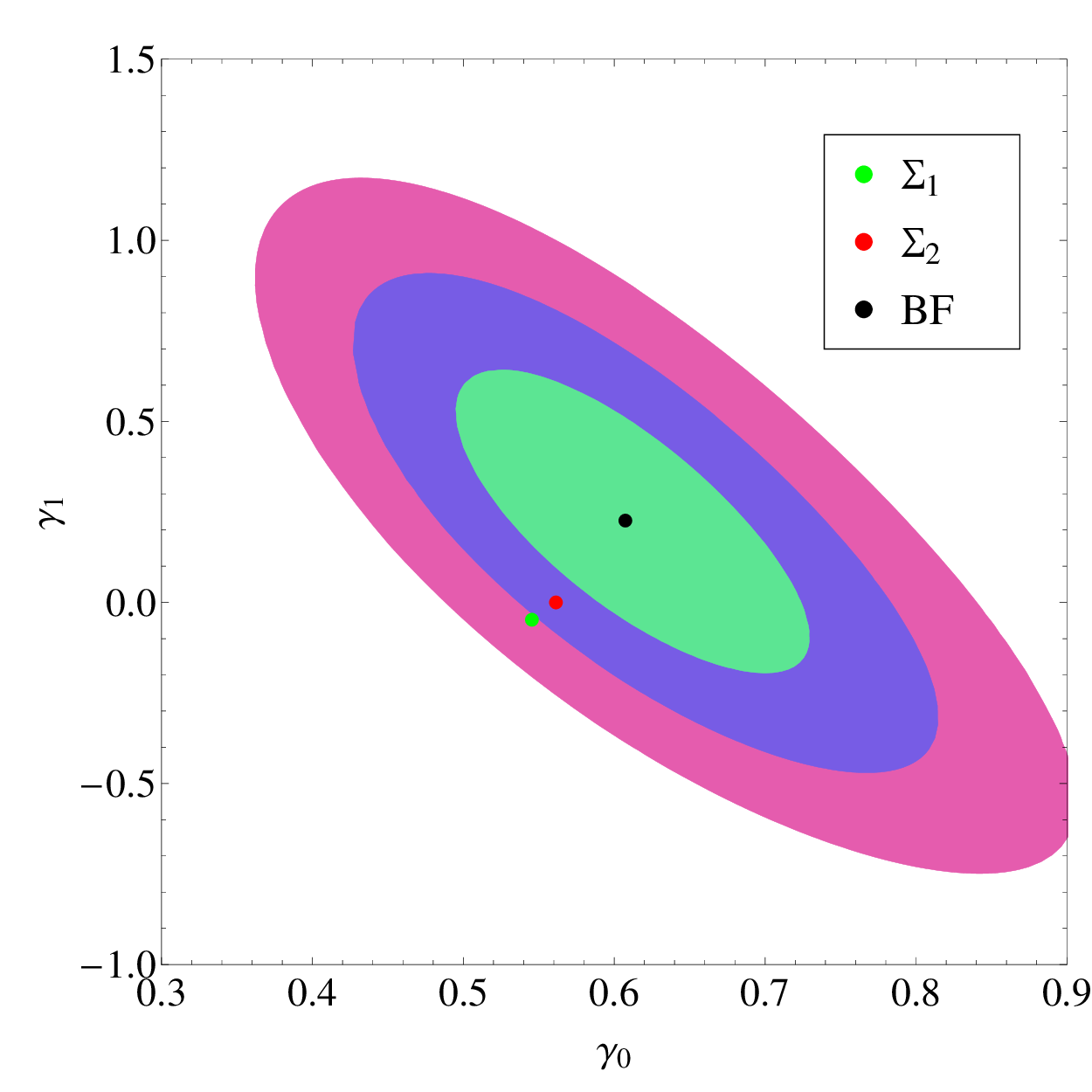}}}
\caption{The $\Lambda$CDM {\em expansion model:} Likelihood contours (for $\Delta \chi^{2}=-2{\rm ln}{\cal L}/{\cal L}_{\rm max}$ equal to 2.30, 6.18
and 11.83, corresponding to 1$\sigma$, 2$\sigma$ and $3\sigma$ confidence levels) in the $(\gamma_{0},\gamma_{1})$. The top row shows the contours when the rest of the parameters are fixed to their best-fit values, while the bottom row when they are marginalized over, while the left and right panels show the
results based on the $\Gamma_{1-2}$ parametrizations respectively. We also include the theoretical $\Lambda$CDM $(\gamma_{0},\gamma_{1})$ values given in section 3, $\Sigma_1=\left(6/11,\gamma_1(6/11,\Omega_{m0,bf})\right)$ and $\Sigma_2=\left(\gamma_{0,bf},\gamma_1(\gamma_{0,bf},\Omega_{m0,bf})\right)$. \label{fig:contours}}
\end{figure*}

\begin{figure*}[t!]
\centering
\vspace{0cm}\rotatebox{0}{\vspace{0cm}\hspace{0cm}\resizebox{0.45\textwidth}{!}{\includegraphics{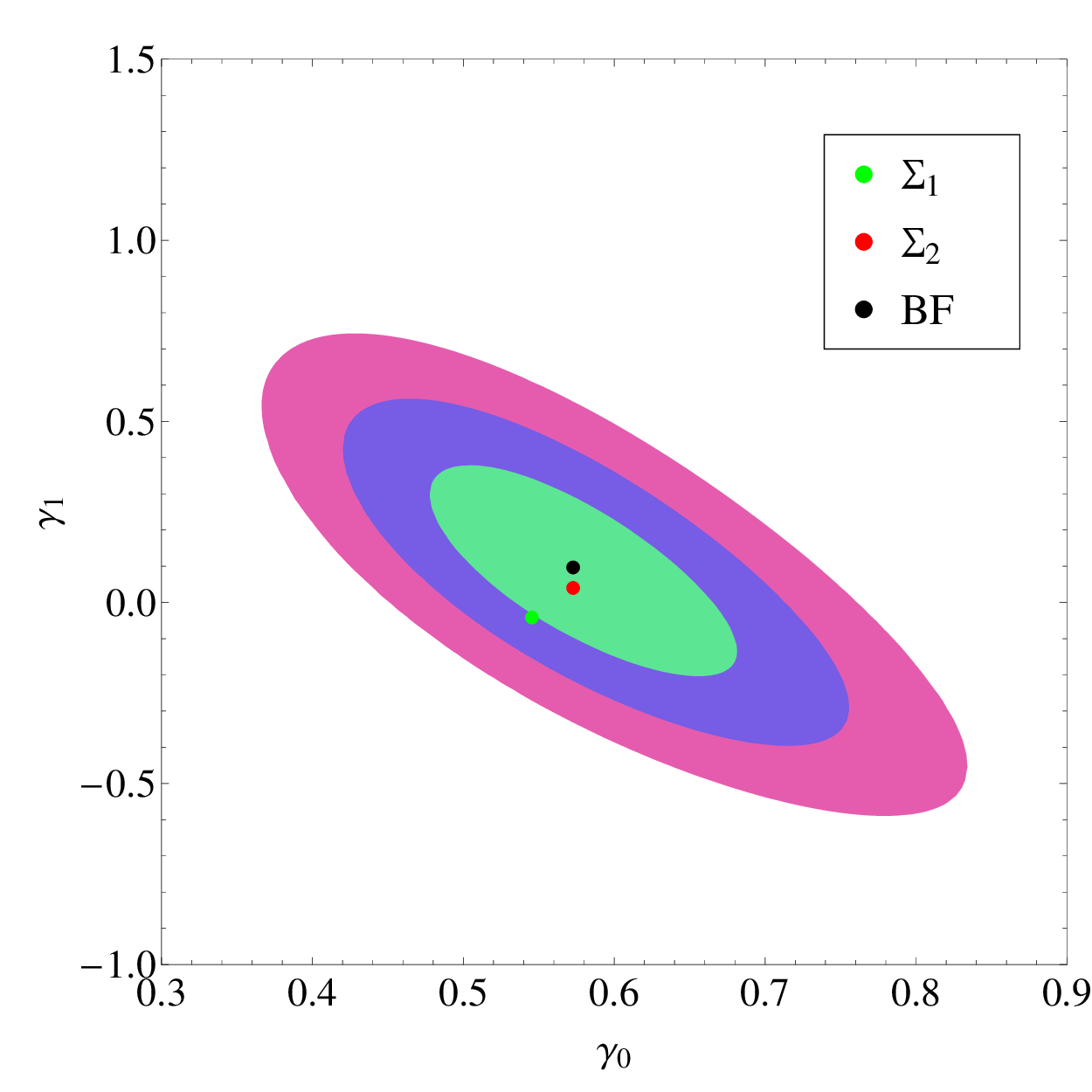}}}
\vspace{0cm}\rotatebox{0}{\vspace{0cm}\hspace{0cm}\resizebox{0.45\textwidth}{!}{\includegraphics{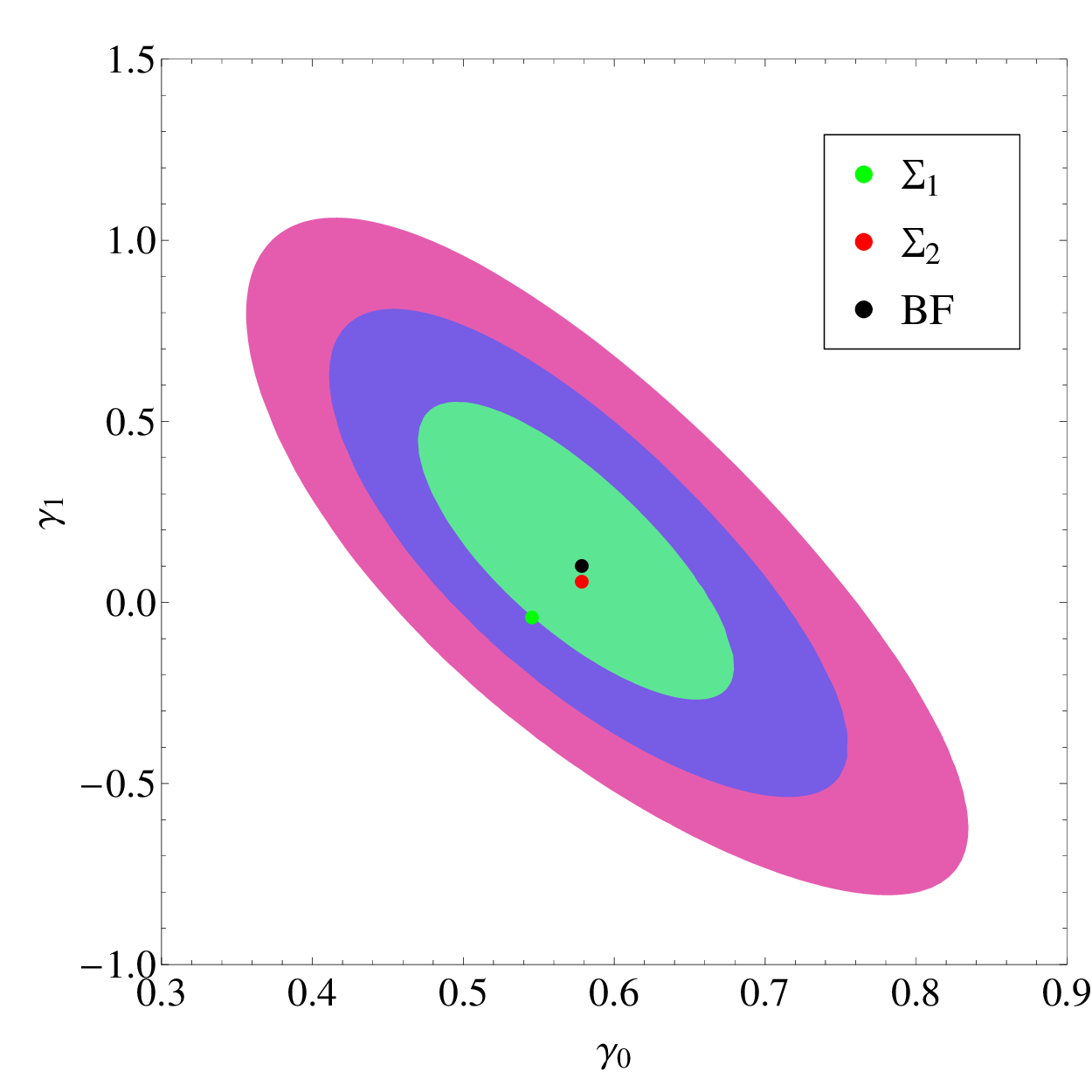}}}
\caption{The Likelihood contours for the $f_{1}$CDM expansion model (for more definitions see caption of Fig.\ref{fig:contours}). Here the colored dots correspond to the theoretical $f_{1}$CDM $(\gamma_{0},\gamma_{1})$ pair provided in section 3A, $\Sigma_1=\left(6/11,\gamma_1(6/11,\Omega_{m0,bf})\right)$ and $\Sigma_2=\left(\gamma_{0,bf},\gamma_1(\gamma_{0,bf},\Omega_{m0,bf})\right)$. \label{fig:contours1}}
\end{figure*}

\begin{figure*}[t!]
\centering
\vspace{0cm}\rotatebox{0}{\vspace{0cm}\hspace{0cm}\resizebox{0.45\textwidth}{!}{\includegraphics{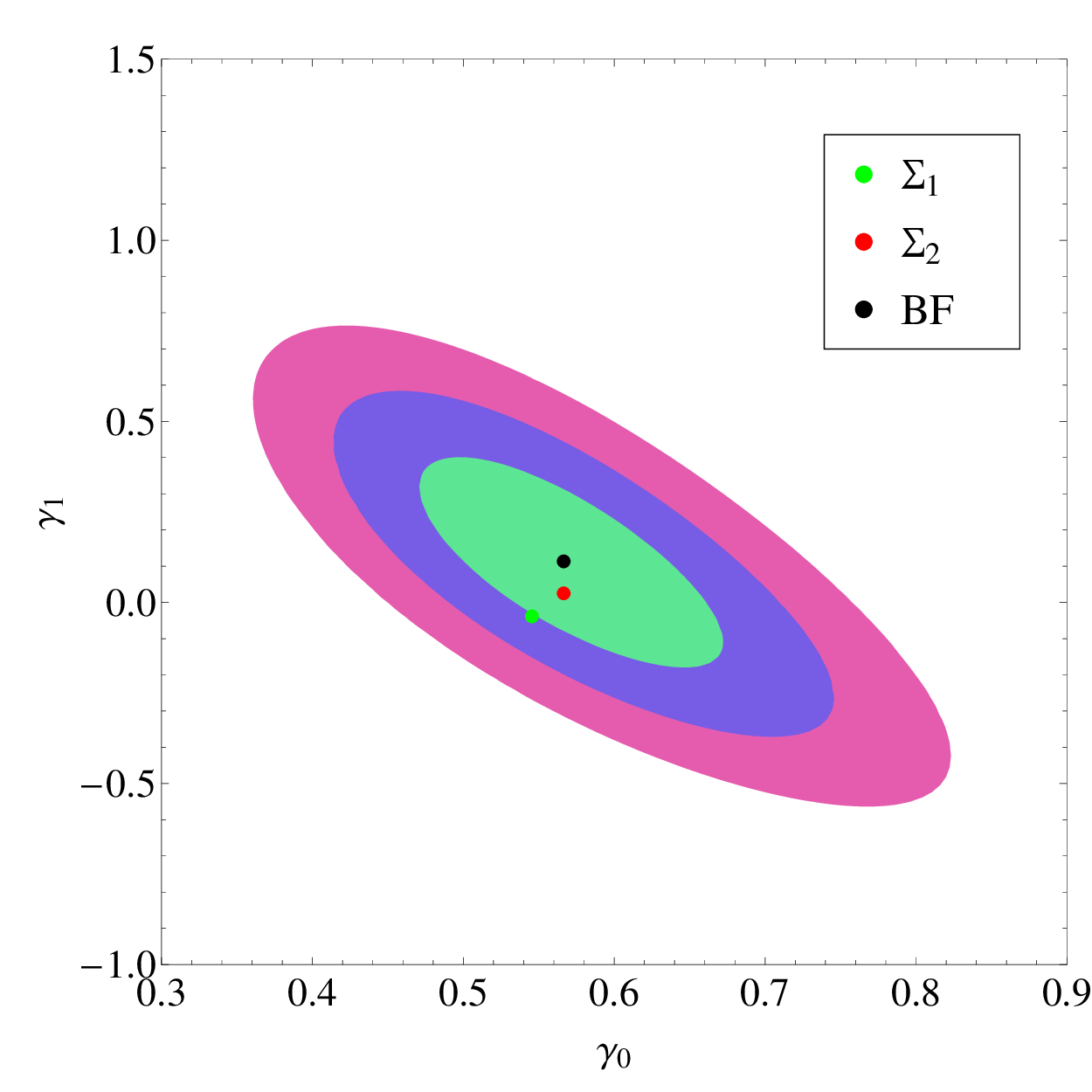}}}
\vspace{0cm}\rotatebox{0}{\vspace{0cm}\hspace{0cm}\resizebox{0.45\textwidth}{!}{\includegraphics{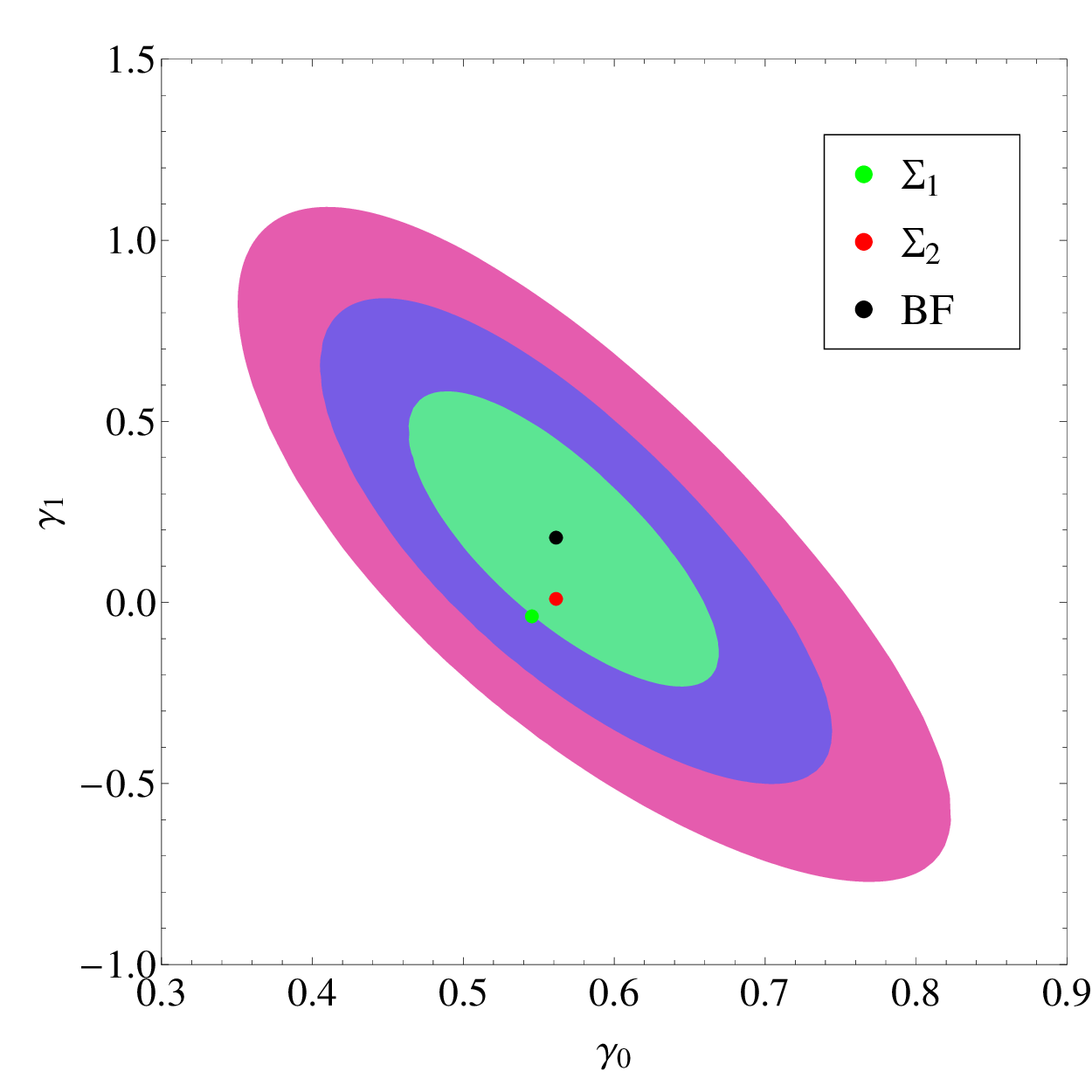}}}
\caption{The Likelihood contours for the $f_{2}$CDM expansion model (for more definitions see caption of Fig.\ref{fig:contours}). Here the colored dots correspond to the theoretical $f_{2}$CDM $(\gamma_{0},\gamma_{1})$ pair provided in section 3A, $\Sigma_1=\left(6/11,\gamma_1(6/11,\Omega_{m0,bf})\right)$ and $\Sigma_2=\left(\gamma_{0,bf},\gamma_1(\gamma_{0,bf},\Omega_{m0,bf})\right)$. \label{fig:contours2}}
\end{figure*}

\begin{figure*}[t!]
\centering
\vspace{0cm}\rotatebox{0}{\vspace{0cm}\hspace{0cm}\resizebox{0.70\textwidth}{!}{\includegraphics{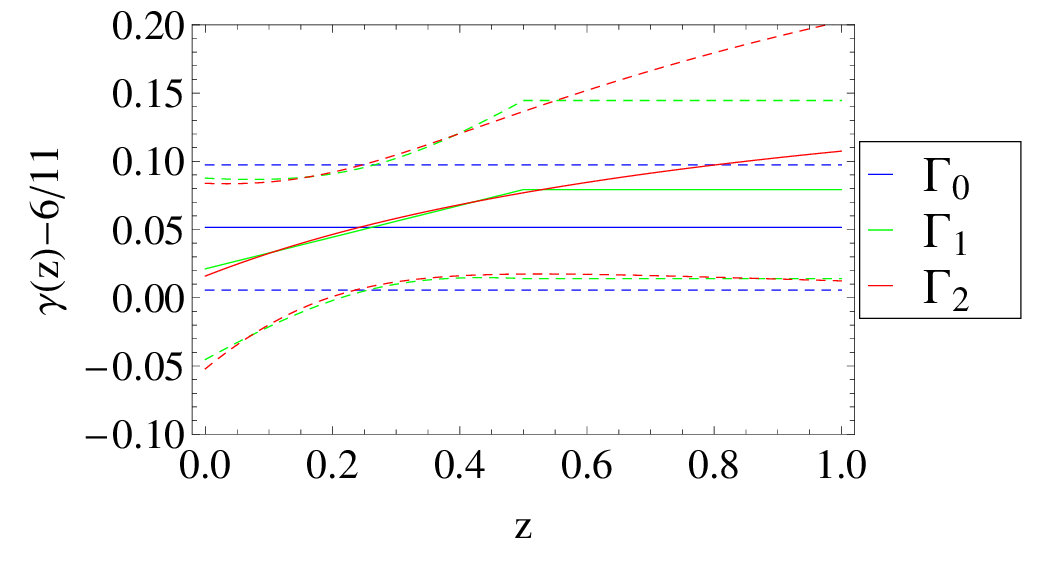}}}
\vspace{0cm}\rotatebox{0}{\vspace{0cm}\hspace{0cm}\resizebox{0.70\textwidth}{!}{\includegraphics{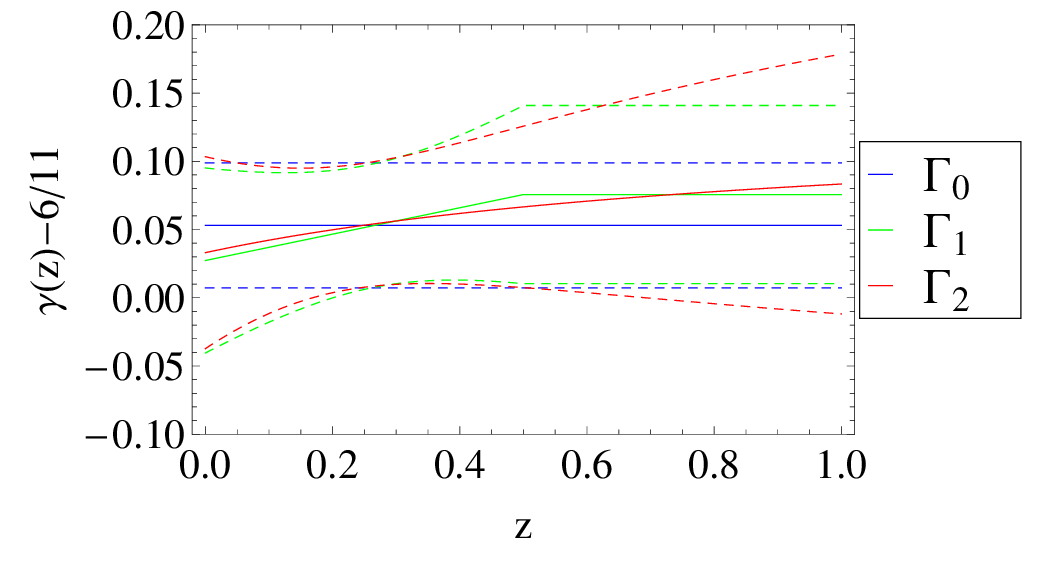}}}
\vspace{0cm}\rotatebox{0}{\vspace{0cm}\hspace{0cm}\resizebox{0.70\textwidth}{!}{\includegraphics{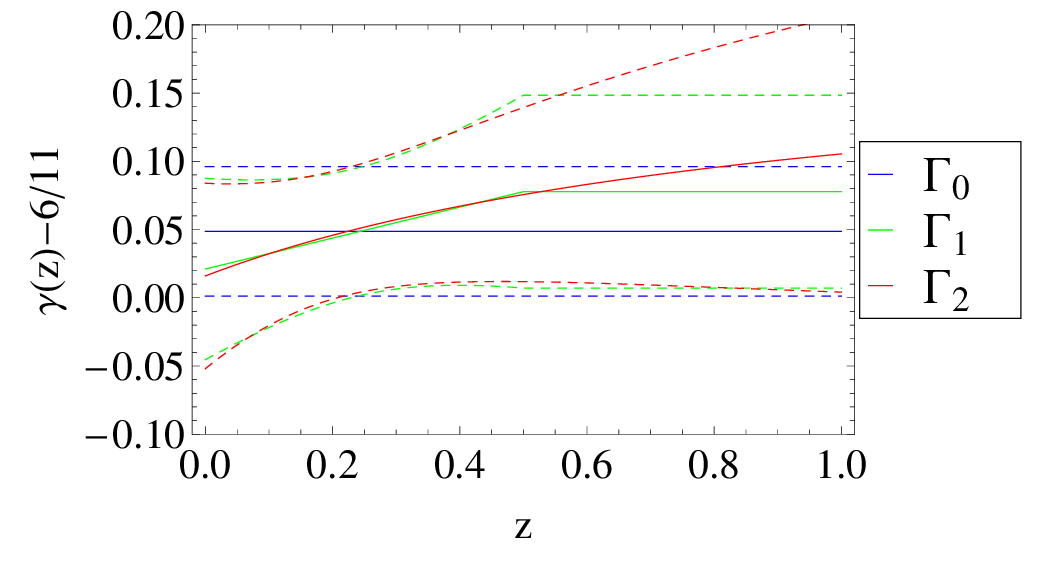}}}
\caption{{\it Top:} The evolution of the growth index
$\gamma(z)-\frac{6}{11}$ for the $\Lambda$CDM model. The lines correspond
to $\Gamma_{0}$ (blue), $\Gamma_{1}$ (green), and $\Gamma_{2}$ (red). {\it
Middle and Bottom:} The evolution of the growth index for the $f_{1}$CDM
and $f_{2}$CDM cosmological models respectively. The lines correspond to
$\Gamma_{0}$ (blue), $\Gamma_{1}$ (green), and $\Gamma_{2}$ (red). The
dashed lines correspond to the $1\sigma$ errors. \label{fig:growth}}
\end{figure*}

\begin{figure*}[t!]
\centering
\vspace{0cm}\rotatebox{0}{\vspace{0cm}\hspace{0cm}\resizebox{1\textwidth}{!}{\includegraphics{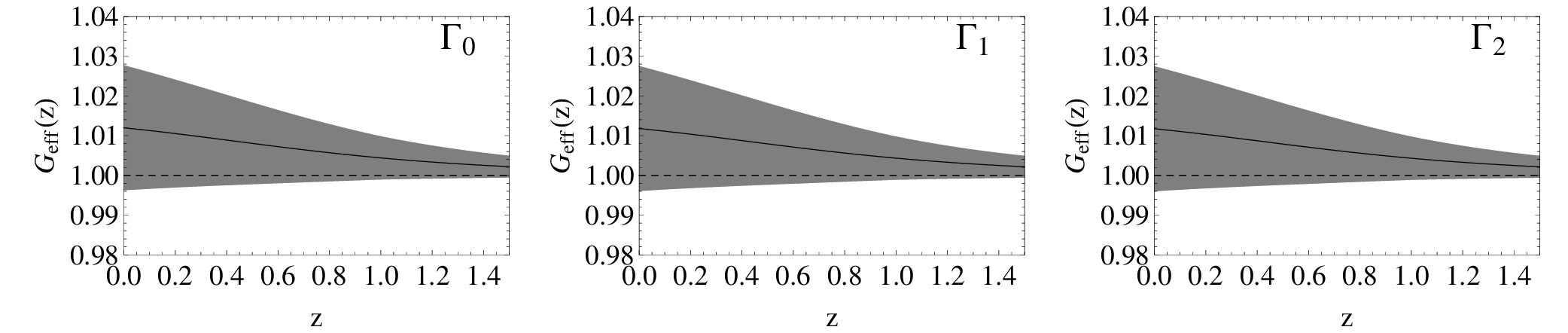}}}
\vspace{0cm}\rotatebox{0}{\vspace{0cm}\hspace{0cm}\resizebox{1\textwidth}{!}{\includegraphics{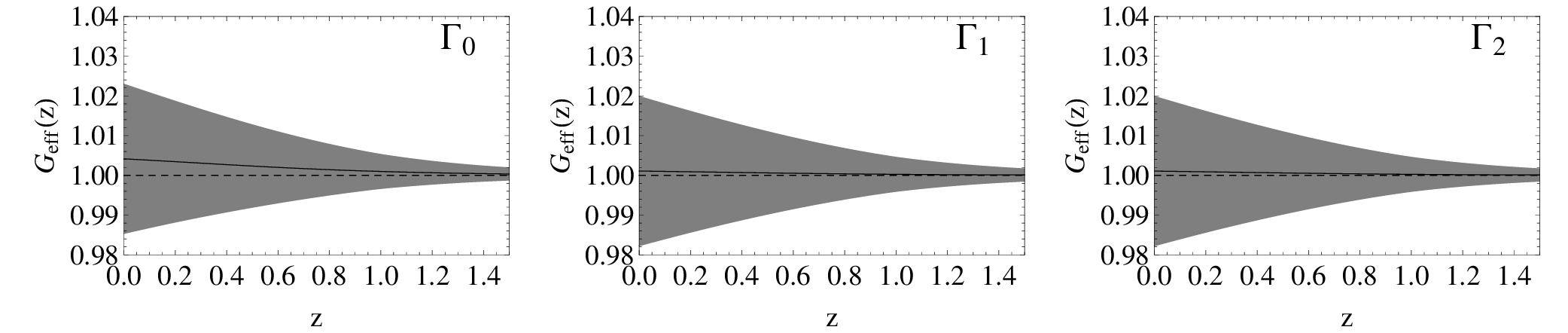}}}
\caption{The evolution of the $G_{eff}(z)$ for the two $f(R)$ models
considered in the text, $f_{1}$CDM (top) and $f_{2}$CDM (bottom), for all
three growth rate parametrizations $\Gamma_{0}$ (left), $\Gamma_{1}$ (middle), and
$\Gamma_{2}$ (right). In all cases we assume for $\kappa=0.1h$Mpc$^{-1}$. \label{fig:geff}}
\end{figure*}

\begin{figure*}[t!]
\centering
\vspace{0cm}\rotatebox{0}{\vspace{0cm}\hspace{0cm}\resizebox{1\textwidth}{!}{\includegraphics{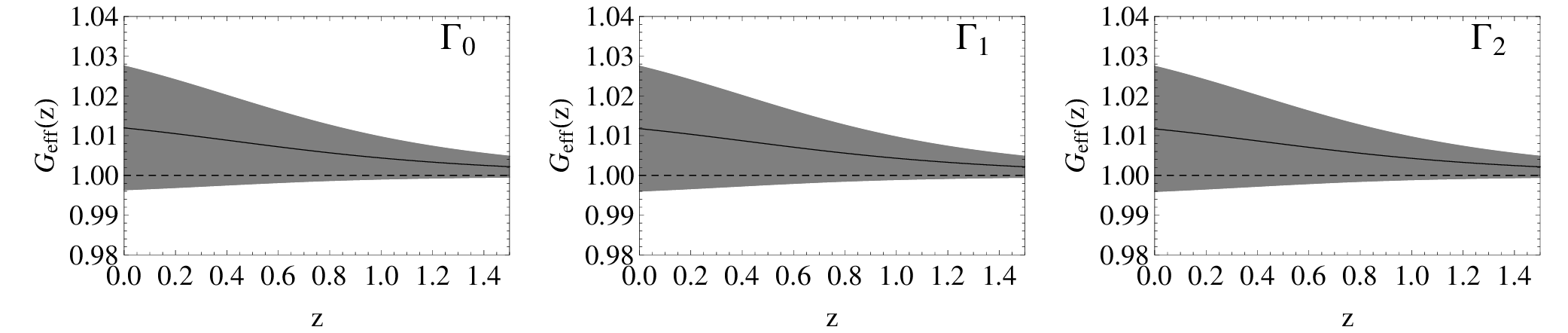}}}
\caption{The evolution of the $G_{eff}(\kappa,z)$ for the HS $f(R)$ model
for $\kappa=0.01h$Mpc$^{-1}$. The plot is practically indistinguishable from Fig. \ref{fig:geff} (top). Therefore, we conclude that for these theories the particular choice of $\kappa$ leaves the results unaffected. \label{fig:geff001}}
\end{figure*}

Concerning the $f_{1}$CDM model (see the middle panel of Fig. \ref{figcon1})
the best fit parameter is $\gamma_0=0.598 \pm 0.046$ with
$\chi^{2}_{min} \simeq 573.855$. Also, we checked the case where $n=2$, compare the middle plot of Fig. \ref{figcon1} with Fig. \ref{figcon1n2}, and we found that our results remain mostly unaffected. Thus throughout the rest of the paper we adopt $n=1$. Also in the case of Starobinsky's $f_{2}$CDM model (see Fig. \ref{figcon1}) the best fit parameter is
$\gamma_0=0.594 \pm 0.047$ with $\chi^{2}_{min} \simeq 574.178$.
In Figure \ref{growthrate}, we plot the measured $\fs_{obs}(z)$
with the estimated growth rate function, $\fs(z)=F(z)\sigma_{8}(z)$.
The value of AIC$_{\Lambda}$($\simeq578.227$) is smaller than the
corresponding $f(R)$ values which indicates that the
$\Lambda$CDM model ($\gamma_{\Lambda}=0.597$) appears now to
fit better than the $f_1$CDM and $f_2$CDM gravity models the expansion and
the growth data. The $|\Delta {\rm AIC}|$=$|{\rm AIC}_{\Lambda}-{\rm AIC}_{f_{1-2}(R)}|$ values (ie., $\sim 2$) indicate that
the growth data are consistent with the $f_1$CDM and $f_2$CDM gravity models with a constant growth index.

At this point we should mention that in all our contour plots, eg Fig. \ref{figcon1}, we have fixed the various parameters to the best-fit values instead if marginalizing over them. It is well known in the literature, see for example \cite{Akrami:2010cz}, that marginalizing over the nuisance parameters and fixing the parameter in general gives different effects since in the former case due to the integration of the likelihood points away from the best-fit and well within the tail of the distribution will contribute in the contour, thus creating a sort of a volume effect.

\subsubsection{Time varying growth index}
Now we concentrate on the $\gamma(z)$ parametrizations, presented
in section 3A. In this case the free parameters of the models
are ${\bf p_{2}} \equiv (\Omega_{m0},\gamma_{0},\gamma_{1})$ and
${\bf p_{2}} \equiv (\Omega_{m0},b,\gamma_{0},\gamma_{1})$ for
the $\Lambda$CDM and $f_1$CDM, $f_2$CDM expansion models respectively.

In Figure \ref{fig:contours} ($\Lambda$CDM model), Figure \ref{fig:contours1}
($f_{1}$CDM model) and Figure \ref{fig:contours2} ($f_{2}$CDM model)
we present the results of our statistical analysis for the $\Gamma_{1}$ (left panel) and $\Gamma_{2}$ (right panel) parametrizations in the $(\gamma_{0},\gamma_{1})$ plane in which the corresponding contours are plotted for 1$\sigma$, 2$\sigma$ and 3$\sigma$ confidence levels. The theoretical $(\gamma_{0},\gamma_{1})$ values (see section 3A) in the $\Lambda$CDM and $f(R)$ expansion models indicated by the colored dots. Overall, we find that the predicted $\Lambda$CDM, $f_{1}$CDM and $f_{2}$CDM
$(\gamma_{0},\gamma_{1})$ solutions of the $\Gamma_{1-2}$
parametrizations are close to the $1\sigma$
borders ($\Delta \chi_{1\sigma}^{2}\simeq 2.30$; see green sectors in Figs.~\ref{fig:contours}, \ref{fig:contours1} and \ref{fig:contours2}).

Furthermore, we also show the corresponding $\gamma_{0}-\gamma_{1}$ contours by marginalizing over the other parameters, see the bottom row in Fig. \ref{fig:contours}, and we find that both procedures are in good agreement within $\sim 1-1.5\sigma$. Statistically this means that the likelihood function is close to being a Gaussian.

Below we briefly discuss the main statistical results:
\noindent
(a) $\Gamma_{1}$ parametrization:
For the usual $\Lambda$ cosmology the likelihood function
peaks at $\gamma_{0}=0.567\pm  0.066$ and
$\gamma_{1}=0.116\pm 0.191$
with $\chi_{min}^{2} \simeq 573.861$. The latter results are in agreement
with previous studies \cite{Port08,Nes08,Dos10,Por}.
In the case of the $f_{1}$CDM and $f_{2}$CDM gravity models
we find that
$(\gamma_{0},\gamma_{1})=(0.573\pm 0.068,0.097\pm 0.195$)
with $\chi_{min}^{2} \simeq 573.633$
and $(\gamma_{0},\gamma_{1})=(0.567\pm 0.066,0.113\pm 0.199$) and
with $\chi_{min}^{2} \simeq 573.857$ respectively.

\noindent
(b) $\Gamma_{2}$ parametrization:
The best fit values are: (i) for $\Lambda$CDM we have
$\gamma_{0}=0.561\pm0.068$,
$\gamma_{1}=0.183\pm0.269$ ($\chi_{min}^{2} \simeq 573.767$),
(ii) in the case of $f_{1}$CDM model we obtain
$\gamma_{0}=0.579\pm0.070$,
$\gamma_{1}=0.101\pm0.275$ ($\chi_{min}^{2} \simeq 573.585$)
and (iii) for $f_{2}$CDM gravity model
we find
$\gamma_{0}=0.561\pm0.068$,
$\gamma_{1}=0.179\pm0.279$ ($\chi_{min}^{2} \simeq 573.765$).

Finally, as we have already mentioned in Table \ref{tab:growth1}, one may see a
more compact presentation of our statistical results.
In Figure \ref{fig:growth} we present the evolution of the growth index
for various parametrizations.
In the all three cases of the concordance $\Lambda$ and
$f_{2}$CDM cosmological models (see upper and bottom panels of Fig.\ref{fig:growth}) the relative growth index difference of the various fitted $\gamma(z)$
models indicates that the $\Gamma_{1-2}$ parametrizations have a very similar
redshift dependence for $z \le 0.5$, while the $\Gamma_{2}$ parametrization
shows very large such deviations for $z>0.5$. Based on the $f_{1}$CDM gravity model
(middle panel of Fig.\ref{fig:growth}) we observe that the $\Gamma_{1-2}$ parametrizations provide a similar evolution of the growth index.

Furthermore, in Fig.~\ref{fig:geff} we show the evolution of $G_{eff}(z)$ for
the two $f(R)$ models considered in the text, $f_{1}$CDM (left) and
$f_{2}$CDM (right). The lines correspond to $\Gamma_{0}$ (blue),
$\Gamma_{1}$ (green), and $\Gamma_{2}$ (red). As it can be seen, both
cases predict little evolution for $G_{eff}(z)$ at late times, around
$1.2\%$ for $f_1$CDM and $~0.5\%$ for $f_2$CDM. Furthermore, while
$G_{eff}(z)$ shows almost the same evolution for all three
parameterizations $\Gamma_{0,1,2}$ of $f_1$CDM, this is not the case for
$f_2$CDM where $\Gamma_{0}$ differs significantly from the other two.

Finally, we repeated our analysis by treating $\sigma_8$ as a free
parameter and the corresponding results are in good agreement with our previous analysis within $1\sigma$ with the results of Table \ref{tab:growth1}, thus justifying our choice to fix $\sigma_8$. Specifically, we found:

In the case of the $\Lambda$CDM:
\begin{itemize}
  \item for the $\Gamma_0$ model: $\chi^2=573.254$, $\Omega_{m0}= 0.272 \pm 0.003$, $\gamma_0=0.523 \pm 0.0858$, $\sigma_8= 0.761 \pm 0.038$.
  \item for the $\Gamma_1$ model: $\chi^2=572.618$, $\Omega_{m0}= 0.272\pm0.003$, $\gamma_0 = 0.485\pm0.098$, $\gamma_1 = -0.398\pm0.502$, $\sigma_8 = 0.694\pm0.087$.
  \item for the $\Gamma_2$ model: $\chi^2=572.652$, $\Omega_{m0}= 0.272\pm 0.003$, $\gamma_0 = 0.483\pm0.097$, $\gamma_1 = -0.633\pm0.815$, $\sigma_8 = 0.685\pm0.097$.
\end{itemize}

In the case of the $f_1$CDM:
\begin{itemize}
  \item for the $\Gamma_0$ model: $\chi^2=573.128$, $\Omega_{m0}= 0.271\pm0.003$, $b = 0.104\pm0.142$, $\gamma_0 = 0.533\pm 0.089$, $\sigma_8 = 0.766\pm 0.039$.
  \item for the $\Gamma_1$ model: $\chi^2=573.277$, $\Omega_{m0}= 0.272\pm0.003$, $b = 0.086\pm0.143$, $\gamma_0 = 0.526\pm0.104$, $\gamma_1 = 0.054\pm0.507$, $\sigma_8 = 0.771\pm0.104$.
  \item for the $\Gamma_2$ model: $\chi^2=572.452$, $\Omega_{m0}= 0.272\pm0.003$, $b = 0.072\pm0.144$, $\gamma_0 = 0.490\pm 0.101$, $\gamma_1 = -0.679\pm0.835$, $\sigma_8 = 0.682\pm 0.099$.
\end{itemize}

In the case of the $f_2$CDM:
\begin{itemize}
  \item for the $\Gamma_0$ model: $\chi^2=573.766$, $\Omega_{m0}= 0.272\pm0.004$, $b = 0.109\pm1.48$, $\gamma_0 = 0.585\pm0.091$, $\sigma_8 = 0.786\pm0.041$.
  \item for the $\Gamma_1$ model: $\chi^2=573.693$, $\Omega_{m0}= 0.272\pm0.002$, $b = 0.092\pm1.52$, $\gamma_0 = 0.549\pm0.108$, $\gamma_1 = 0.0934\pm0.551$, $\sigma_8 = 0.790\pm 0.116$.
  \item for the $\Gamma_2$ model: $\chi^2=572.653$, $\Omega_{m0}= 0.272\pm0.004$, $b = 0.082\pm1.71$, $\gamma_0 = 0.483\pm0.097$, $\gamma_1 = -0.635\pm0.408$, $\sigma_8 = 0.684\pm0.038$.
\end{itemize}

\section{Conclusions}
It is well known that the growth index $\gamma$ plays a key role in
cosmological studies because it can be used as a useful tool in order
to test Einstein's general relativity on cosmological scales.
In this article, we utilized the recent growth rate data
provided by the clustering, measured mainly from the PSCz, 2dF, VVDS,
SDSS, 6dF, 2MASS, BOSS and {\em WiggleZ}
galaxy surveys, in order to constrain the growth index.
In particular, performing simultaneously a likelihood analysis of the
recent expansion data (SnIa, CMB shift parameter and BAOs)
together with the growth rate of structure data, in order to
determine the cosmological and the free parameters of the $\gamma(z)$ parametrizations and thus to statistically quantify the ability of $\gamma(z)$
to represent the observations. We consider the following growth index parametrization $\gamma(z)=\gamma_{0}+\gamma_{1}y(z)$
[where $y(z)\equiv 0$, $y(z)=z$ and $1-a(z)$].
Overall, considering a $\Lambda$CDM expansion model
we found that the observed growth index is in agreement, within
$1\sigma$ errors, with the theoretically predicted value of $\gamma_{\Lambda}\simeq 6/11$. Additionally, based on the Akaike information criterion
we shown that for any type of $\gamma(z)$ the combined analysis of the
cosmological (expansion+growth) data can accommodate the
Hu-Sawicky and Starobinsky $f(R)$ gravity models for small values of the deviation parameter $b$.

{\bf Numerical Analysis Files:} The Mathematica and data files used for the numerical analysis of this study may be downloaded from: http://leandros.physics.uoi.gr/fr-constraints/probes.htm .

\section*{Acknowledgements}
We would like to thank J.~Garc\'ia-Bellido, C.~Blake, C.H.~Chuang and I.~Sawicki for very useful and enlightening discussions. S.N. acknowledges financial support from the Madrid Regional Government (CAM) under the program HEPHACOS S2009/ESP-1473-02, from MICINN under grant AYA2009-13936-C06-06 and Consolider-Ingenio 2010 PAU (CSD2007-00060), as well as from the European Union Marie Curie Initial Training Network UNILHC PITN-GA-2009-237920. This research has been co-financed by the European Union (European Social Fund - ESF) and Greek national funds through the Operational Program "Education and Lifelong Learning" of the National Strategic Reference Framework (NSRF) - Research Funding Program: THALIS. Investing in the society of knowledge through the European Social Fund.
SB acknowledges support by the Research Center for
Astronomy of the Academy of Athens
in the context of the program {\it ''Tracing the Cosmic Acceleration''}.

\appendix
\section{Useful formulae\label{apps}}
Here we provide the exact expressions of the coefficients $\delta H_i^2(N)$ for the HS and Starobinsky models for $n=1$. In all cases $H_{\Lambda}^2(N)$ is given by Eq.~(\ref{LCDM1}). For the HS model the first two terms are:
\begin{widetext}
\bear
\frac{\delta H_1^2(N)}{H_0^2}&=&-\frac{H_0^2\left(-1+\Omega _{m0}+\Omega _{r0}\right)^2 \left(6 H_{\Lambda }(N)^2+4 H_{\Lambda }'(N)^2+H_{\Lambda }(N) \left(15 H_{\Lambda }'(N)+2 H_{\Lambda }''(N)\right)\right)}{2 H_{\Lambda }(N) \left(2 H_{\Lambda }(N)+H_{\Lambda }'(N)\right)^3} \label{HSap1} \\
\frac{\delta H_2^2(N)}{H_0^2}&=& -(H_0^4 (-1+\Omega _m+\Omega _r){}^3 (128 H_{\Lambda }(N){}^8-32 H_0^2 (-1+\Omega _m+\Omega _r) H_{\Lambda }'(N){}^6+32 H_{\Lambda }(N){}^7 (25 H_{\Lambda }'(N)\nn \\&&+3 H_{\Lambda }''(N))-2 H_0^2 (-1+\Omega _m+\Omega _r) H_{\Lambda }(N) H_{\Lambda }'(N){}^4 (139 H_{\Lambda }'(N)+22 H_{\Lambda }''(N))\nn \\&&+16 H_{\Lambda }(N){}^6 (9 H_0^2 (-1+\Omega _m+\Omega _r)+89 H_{\Lambda }'(N){}^2+12 H_{\Lambda }'(N) H_{\Lambda }''(N))\nn \\&&+H_{\Lambda }(N){}^2 H_{\Lambda }'(N){}^2 (-749 H_0^2 (-1+\Omega _m+\Omega _r) H_{\Lambda }'(N){}^2+9 H_{\Lambda }'(N){}^4-48 H_0^2 (-1+\Omega _m+\Omega _r) H_{\Lambda }''(N){}^2\nn \\&&-4 H_0^2 (-1+\Omega _m+\Omega _r) H_{\Lambda }'(N) (74 H_{\Lambda }''(N)-3 H_{\Lambda }{}^{(3)}(N)))+8 H_{\Lambda }(N){}^5 (144 H_0^2 (-1+\Omega _m+\Omega _r) H_{\Lambda }'(N)\nn \\&&+146 H_{\Lambda }'(N){}^3+18 H_{\Lambda }'(N){}^2 H_{\Lambda }''(N)+H_0^2 (-1+\Omega _m+\Omega _r) (15 H_{\Lambda }''(N)-6 H_{\Lambda }{}^{(3)}(N)-H_{\Lambda }{}^{(4)}(N)))\nn \\&&+4 H_{\Lambda }(N){}^4 (540 H_0^2 (-1+\Omega _m+\Omega _r) H_{\Lambda }'(N){}^2+124 H_{\Lambda }'(N){}^4+12 H_{\Lambda }'(N){}^3 H_{\Lambda }''(N)\nn \\&&+3 H_0^2 (-1+\Omega _m+\Omega _r) H_{\Lambda }''(N) (17 H_{\Lambda }''(N)+4 H_{\Lambda }{}^{(3)}(N))+2 H_0^2 (-1+\Omega _m+\Omega _r) H_{\Lambda }'(N) (129 H_{\Lambda }''(N)\nn \\&&+12 H_{\Lambda }{}^{(3)}(N)-H_{\Lambda }{}^{(4)}(N)))-2 H_{\Lambda }(N){}^3 (84 H_0^2 (-1+\Omega _m+\Omega _r) H_{\Lambda }'(N){}^3-53 H_{\Lambda }'(N){}^5\nn \\&&-3 H_{\Lambda }'(N){}^4 H_{\Lambda }''(N)+21 H_0^2 (-1+\Omega _m+\Omega _r) H_{\Lambda }''(N){}^3+3 H_0^2 (-1+\Omega _m+\Omega _r) H_{\Lambda }'(N) H_{\Lambda }''(N) \times \nn \\&&(41 H_{\Lambda }''(N)-4 H_{\Lambda }{}^{(3)}(N))+H_0^2 (-1+\Omega _m+\Omega _r) H_{\Lambda }'(N){}^2 (217 H_{\Lambda }''(N)-42 H_{\Lambda }{}^{(3)}(N)+H_{\Lambda }{}^{(4)}(N)))))\nn \\&&/(4 H_{\Lambda }(N){}^4 (2 H_{\Lambda }(N)+H_{\Lambda }'(N)){}^8) \label{HSap2}
\eear

while for the Starobinsky model the first two terms are:

\be
\frac{\delta H_2^2(N)}{H_0^2}=\frac{H_0^4\left(-1+\Omega _{m0}+\Omega _{r0}\right){}^3 \left(8 H_{\Lambda }(N){}^2+9 H_{\Lambda }'(N){}^2+H_{\Lambda }(N) \left(34 H_{\Lambda }'(N)+6 H_{\Lambda }''(N)\right)\right)}{4 H_{\Lambda }(N){}^2 \left(2 H_{\Lambda }(N)+H_{\Lambda }'(N)\right){}^4} \label{Starap1}
\ee

\bear
&&\frac{\delta H_4^2(N)}{H_0^2}= -H_0^{8}\left(-1+\Omega _{m0}+\Omega _{r0}\right){}^5 (192 H_{\Lambda }(N){}^8-486 H_0^2\left(-1+\Omega _{m0}+\Omega _{r0}\right) H_{\Lambda }'(N){}^6+320 H_{\Lambda }(N){}^7 (6 H_{\Lambda }'(N)-\nn \\
&&+H_{\Lambda }''(N))-
18 H_0^2\left(-1+\Omega _{m0}+\Omega _{r0}\right) H_{\Lambda }(N) H_{\Lambda }'(N){}^4 \left(263 H_{\Lambda }'(N)+48 H_{\Lambda }''(N)\right)+16 H_{\Lambda }(N){}^6 (32 \left(-1+\Omega _{m0}+\Omega _{r0}\right)+\nn \\&&
5 H_{\Lambda }'(N) \left(47 H_{\Lambda }'(N)+8 H_{\Lambda }''(N)\right))+H_{\Lambda }(N){}^2 H_{\Lambda }'(N){}^2 (-15562 H_0^2\left(-1+\Omega _{m0}+\Omega _{r0}\right) H_{\Lambda }'(N){}^2+25 H_{\Lambda }'(N){}^4-\nn \\&&
972 H_0^2\left(-1+\Omega _{m0}+\Omega _{r0}\right) H_{\Lambda }''(N){}^2-12 H_0^2\left(-1+\Omega _{m0}+\Omega _{r0}\right) H_{\Lambda }'(N) \left(532 H_{\Lambda }''(N)-15 H_{\Lambda }{}^{(3)}(N)\right))+\nn \\&&
4 H_{\Lambda }(N){}^4 (4422 H_0^2\left(-1+\Omega _{m0}+\Omega _{r0}\right) H_{\Lambda }'(N){}^2+345 H_{\Lambda }'(N){}^4+40 H_{\Lambda }'(N){}^3 H_{\Lambda }''(N)+\nn \\&&
72 H_0^2\left(-1+\Omega _{m0}+\Omega _{r0}\right) H_{\Lambda }''(N) \left(9 H_{\Lambda }''(N)+2 H_{\Lambda }{}^{(3)}(N)\right)+18 H_0^2\left(-1+\Omega _{m0}+\Omega _{r0}\right) H_{\Lambda }'(N) \times \nn \\&&
\left(175 H_{\Lambda }''(N)+20 H_{\Lambda }{}^{(3)}(N)-H_{\Lambda }{}^{(4)}(N)\right))+2 H_{\Lambda }(N){}^3 (-7000 H_0^2\left(-1+\Omega _{m0}+\Omega _{r0}\right) H_{\Lambda }'(N){}^3+\nn \\&&
148 H_{\Lambda }'(N){}^5+10 H_{\Lambda }'(N){}^4 H_{\Lambda }''(N)-324 H_0^2\left(-1+\Omega _{m0}+\Omega _{r0}\right) H_{\Lambda }''(N){}^3-36 H_0^2\left(-1+\Omega _{m0}+\Omega _{r0}\right) H_{\Lambda }'(N) H_{\Lambda }''(N) \times \nn \\&&
\left(63 H_{\Lambda }''(N)-4 H_{\Lambda }{}^{(3)}(N)\right)-9 H_0^2\left(-1+\Omega _{m0}+\Omega _{r0}\right) H_{\Lambda }'(N){}^2 \left(609 H_{\Lambda }''(N)-66 H_{\Lambda }{}^{(3)}(N)+H_{\Lambda }{}^{(4)}(N)\right))+\nn \\&&
8 H_{\Lambda }(N){}^5 (824 H_0^2\left(-1+\Omega _{m0}+\Omega _{r0}\right) H_{\Lambda }'(N)+400 H_{\Lambda }'(N){}^3+60 H_{\Lambda }'(N){}^2 H_{\Lambda }''(N)+3 H_0^2\left(-1+\Omega _{m0}+\Omega _{r0}\right) \times\nn \\&&
\left(29 H_{\Lambda }''(N)-3 \left(6 H_{\Lambda }{}^{(3)}(N)+H_{\Lambda }{}^{(4)}(N)\right)\right)))/\left(16 H_{\Lambda }(N){}^6 \left(2 H_{\Lambda }(N)+H_{\Lambda }'(N)\right){}^{10}\right) \label{Starap2}
\eear
\end{widetext}

\end{document}